\documentclass[aps,prc,groupedaddress,nofootinbib,twocolumn,preprintnumbers]{revtex4-2}

\usepackage{amsmath,amssymb,bm,mathrsfs}
\usepackage{graphicx,hyperref}
\usepackage{color}
\usepackage[stable]{footmisc}

\hypersetup{
    colorlinks=true,
    linkcolor=red, 
    citecolor=green, 
    urlcolor=blue
}

\begin{document}
\title{Dynamics of the conserved net-baryon density near QCD critical point within inhomogeneous quark-gluon plasma profile}

\author{Shanjin Wu\textsuperscript{1}}
\email{shanjinwu@lzu.edu.cn}
\affiliation{\textsuperscript{1}School of Nuclear Science and Technology, Lanzhou University, Lanzhou 730000, China}

\begin{abstract}
This paper investigates the dynamics of the net-baryon multiplicity fluctuations near the QCD critical point, within the inhomogeneous temperature and baryon chemical potential profile of quark-gluon plasma from the hydrodynamics. The Langevin dynamics of conserved net-baryon density are solved numerically within the temperature and chemical potential profile  borrowing from hydrodynamic simulation. It is found that the local systems at different rapidities reach the critical point at different proper times, owing to the inhomogeneous temperature and chemical potential. As a result, it is observed the pronounced enhancement of the magnitude for the net-baryon multiplicity fluctuations with large rapidity acceptance at the freeze-out surface, which is the consequence of the combined effect of critical slowing down and inhomogeneous profile.
\end{abstract}
\maketitle
\section{Introduction}\label{sec:intro}

Searching the Quantum Chromodynamics (QCD) critical point is one of the most important goals of the relativistic heavy-ion collisions. The transition from the quark-gluon plasma (QGP) phase to the hadron phase is revealed as a crossover at the vanishing baryon chemical potential ($\mu_B\simeq0$) by lattice QCD \cite{Aoki:2006we,Ding:2015ona,Bazavov:2019lgz,Ratti:2018ksb}. While the first-order phase transition together with the critical point is predicted by the effective theories of QCD at finite chemical potential~\cite{Fischer:2018sdj,Fukushima:2010bq,Fukushima:2013rx,Fu:2022gou}. The most important property of the critical point is the long-range correlation and large fluctuations. As a consequence, the non-monotonic behavior is conjectured as the characteristic signature of the QCD phase transition~\cite{Stephanov:2011pb,Athanasiou:2010kw}.

The Beam Energy Scan program at RHIC has been dedicated to exploring the QCD phase structure. The preliminary non-monotonic behavior of net-proton multiplicity fluctuations as a function of colliding energy has been observed~\cite{STAR:2020tga,
STAR:2021iop}, which is consistent with the theoretical prediction~\cite{Stephanov:2011pb,Athanasiou:2010kw}. However, the statistics of the measurement are insufficient to conclude the observation of this non-monotonicity so far, which is looking forward to the higher statistics in the
coming Bean Energy Scan phase two program.

On the other hand, the final confirmation of the existence of the QCD critical point requires the comparison between the experimental measurement and theoretical prediction. After decades of efforts, remarkable progress has been made in the theoretical modeling of the dynamics near the QCD critical point within the complex evolution of the relativistic heavy-ion collisions. Please see e.g., Refs.\cite{Asakawa:2015ybt,Bzdak:2019pkr,Bluhm:2020mpc,Wu:2021xgu,An:2021wof,Du:2024wjm} for recent review. Due to the fast-expanding QGP fireball, the non-equilibrium critical fluctuations become non-trivial compared with the equilibrium ones. For example, the magnitude of the fluctuations is suppressed~\cite{Berdnikov:1999ph,Nonaka:2004pg}, the sign could be reversed~\cite{Mukherjee:2015swa}, and the largest fluctuations not necessarily correspond to the trajectory closest to the critical point~\cite{Tang:2023zvj}. Subsequently, several models have been developed to incorporate the dynamics of the critical effects. For instance, the dynamics of conserved charge were constructed~\cite{Nahrgang:2018afz,Nahrgang:2020yxm,Sakaida:2017rtj,Pihan:2022xcl}and non-monotonic behavior of multiplicity fluctuations with respect to acceptance was predicted~\cite{Sakaida:2017rtj,Pihan:2022xcl}. To investigate the dynamics of the QGP fireball near the QCD critical point, the dynamics of the critical fluctuations were also coupled to the dynamics of the fireball as an additional degree of freedom, such as the stochastic dynamics of order parameter field in non-equilibrium chiral hydrodynamics~\cite{Nahrgang:2011mg,Nahrgang:2011vn,Herold:2014zoa}, the stochastic noise terms in the fluctuating hydrodynamics~\cite{Kapusta:2011gt, An:2019csj,An:2020vri,An:2021wof} and the slow mode in the hydro+~\cite{Stephanov:2017ghc}. 

The dynamics of the conserved quantity are of particular interest because the corresponding diffusion process takes time. Therefore, the correlation between particles far away from each other preserves the early evolution of diffusion~\cite{Sakaida:2017rtj}, and the non-monotonic behavior with respect to acceptance could be regarded as the imprint of the evolving trajectory passing through the critical region. In addition, multi-particle correlation becomes indispensable as the system is extremely close to the critical point. Thus, the dynamics of conserved net-baryon density with higher-order correlation was constructed in the longitudinal Bjorken expansion system with uniform temperature and chemical potential profile~\cite{Pihan:2022xcl}. It was found that the pronounced minimum of kurtosis presents at the intermediate rapidity due to the existence of the critical point.

In the realistic context, the temperature and chemical point are not uniform across the QGP profile because of the baryon stopping effects~\cite{Shen:2017bsr,Du:2023gnv}. As a result, different regions of the QGP profile experience distinct trajectories across the critical region~\cite{Du:2021zqz}, and the rapidity dependence is expected to deform non-trivially~\cite{Brewer:2018abr}. In the dynamics of conserved variables, the realistic QGP profile is particularly important because its rapidity dependence on correlation preserves the history along the evolving trajectory. Therefore, it is essential to investigate the rapidity dependence of the multiplicity fluctuations within the more realistic QGP profile. This work studies the dynamics of the conserved net-baryon density near the critical point within the inhomogeneous temperature and chemical potential QGP profile. With the profile obtained from the hydrodynamic simulation, this works also studies the multiplicity fluctuations at freeze-out hypersurface. It is found that the magnitude of the second-order cumulants and kurtosis enhance significantly at large rapidity, which is the consequence of combined effects of critical slowing down and inhomogeneous QGP profile.

\section{Model and Setups}\label{sec:model}

To study the dynamics of QCD phase transition, one of the natural and simple dynamical models is just focusing on the dynamics of the order parameter field. However, it was pointed out that the dynamics of conserved baryon density is the slowest mode of the system near the critical point when considering the coupling between baryon density and order parameter field~\cite{Son:2004iv}. Therefore, the simplest description of the dynamics near the QCD critical point is only considering the dynamics of conserved baryon density and treating the dynamics of other degrees of freedom as the heat bath.

This study focuses on the $1+1$-dimensional conserved dynamics of net-baryon density $n_B$ near the QCD critical point along the longitudinal direction in Milne frame, i.e. the proper-time $\tau=\sqrt{t^2-z^2}$ and space-time rapidity $\eta=\frac{1}{2} \ln [(t+z)/(t-z)]$: 
\begin{align}\label{eq:diffeq}
    \partial_\tau n_B(\tau,\eta)=\frac{D\chi_2T}{\tau}\frac{\partial^2}{\partial \eta^2} \biggl(\frac{1}{T}\frac{\delta F}{\delta n_B}\biggr)-\partial_\eta \zeta(\tau,\eta),
\end{align}
with the white noise $\zeta(\tau,\eta)$ has only one non-zero correlation:
\begin{align}\label{Eq:noise}
    \langle \zeta(\tau,\eta)\zeta(\tau',\eta')\rangle = \frac{2D\chi_2T}{A\tau}\delta(\tau-\tau')\delta(\eta-\eta').
\end{align}
Here, $D$ is the diffusion coefficient, and $T$ is the temperature. As in Ref.~\cite{Nahrgang:2020yxm}, the transverse area is set as $A=1\mbox{fm}^2$ with the consideration of fluctuation-dissipation relation in the 1+1-dimensional evolution equation Eq.\eqref{eq:diffeq}. The equation of $n_B$ can be derived from the conservation of net-baryon number $\partial_\mu N^\mu=0$ within the Bjorken flow $u^\mu=(\cosh\eta,0,0,\sinh\eta)$. Please see Refs.~\cite{Pihan:2022xcl,Ling:2013ksb} for detail and e.g., Refs.~\cite{Chao:2020kcf,Chao:2023kvz,2024arXiv240315825H} for the further extension.

The effective potential $F$ near the critical point can be parameterized as~\cite{Pihan:2022xcl}: 
\begin{align}\label{Eq:effective_potential}
    F[n_B]=\int d^2x_\perp d\eta \biggl[&\frac{1}{2}\frac{1}{\chi_2 \tau} n_B(\tau,\eta)^2+\frac{1}{2}\frac{K}{\tau^3}\biggl(\frac{\partial}{\partial \eta} n_B(\tau,\eta)\biggr)^2\nonumber\\
    &\quad+\frac{1}{3}\frac{\lambda_3}{\tau^2}n_B(\tau,\eta)^3+\frac{1}{4}\frac{\lambda_4}{\tau^3}n_B(\tau,\eta)^4\biggr],
\end{align}
where $K$ represents the surface tension coefficient and is treated as a constant in this study. The second-order baryon susceptibility $\chi_2$, third- and fourth-order coupling coefficients $\lambda_3,\lambda_4$ contain two parts: regular part estimated from lattice QCD simulation and singular part ($\chi_2^{\mbox{cri}}(T,\mu),\lambda^{\mbox{cri}}_3(T,\mu),\lambda^{\mbox{cri}}_4(T,\mu)$) mapped from three-dimensional Ising model ($\chi_2^{\mbox{cri}}(r,h),\lambda^{\mbox{cri}}_3(r,h),\lambda^{\mbox{cri}}_4(r,h)$) (here $r,h$ are Ising variables). A complete description of the parameterization of the coefficients can be found in  Appendix \ref{app:parametrization}. The effective potential $F$ parameterized in Eq.\eqref{Eq:effective_potential}, including the second to fourth order of the field as well as the surface tension term, is the typical Landau-Ginzburg form employed to study the system near the phase transition. The cubic term is neglected in Ref.~\cite{Pihan:2022xcl} because of the limited knowledge of regular third-order baryon susceptibility. For theoretical consistency, the cubic term is considered in this work with the regular part of the third-order baryon susceptibility estimated from effective models~\cite{Motornenko:2019arp,Mukherjee:2016nhb}. Please see Appendix \ref{app:parametrization} for details.

\begin{figure}[htb]
  \includegraphics[width=0.48\textwidth]{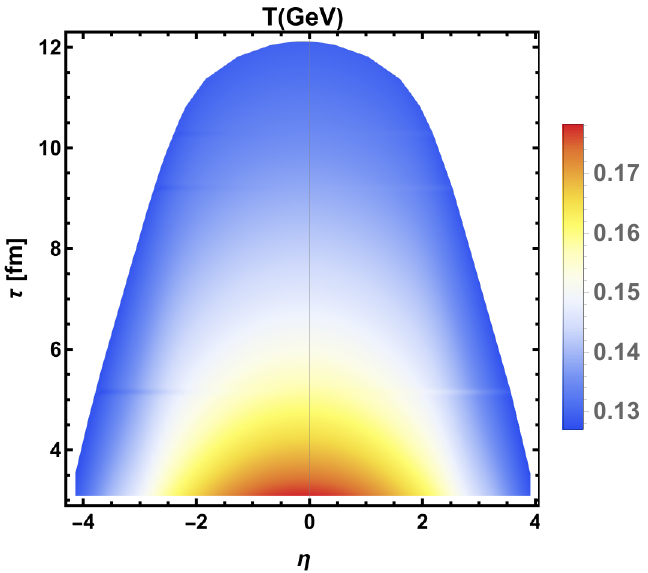}
  \includegraphics[width=0.48\textwidth]{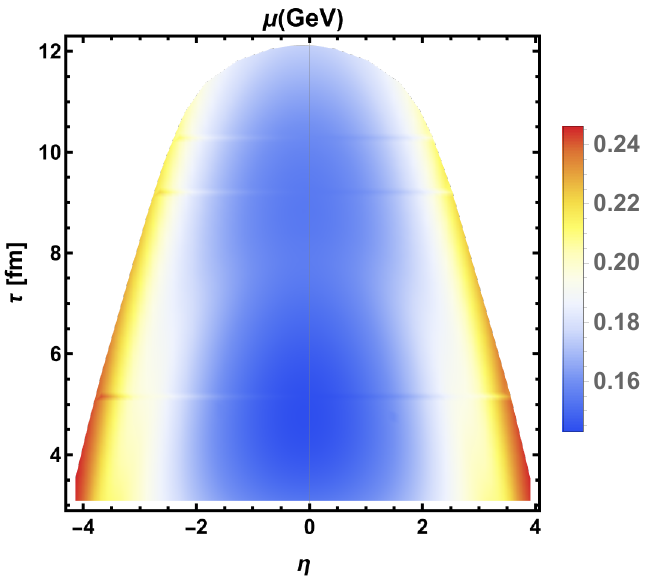}
  \caption{Time evolution of the temperature $T$ (upper panel) and chemical potential $\mu$ (lower panel) profiles in the rapidity $\eta$ space, obtained from the hydrodynamic simulation of Au+Au collisions at 19.6GeV with $0-5\%$ centrality. The $T$ and $\mu$ profiles are depicted from $\tau_0=3$fm, the starting time of Eq.\ref{eq:diffeq}.}
  \label{fig:Tmu}
\end{figure}

The parametrization of the baryon susceptibility and coupling coefficients on the QCD phase diagram requires the time evolution profiles for the QGP fireball: temperature $T(\tau,\eta)$ and chemical potential $\mu(\tau,\eta)$. The time evolution of the QGP profiles is obtained from the 3+1-d hydrodynamic MUSIC simulation~\cite{Shen:2014vra}, as in Ref.\cite{Tang:2023zvj}. The initial profiles constructed from the transport model AMPT~\cite{Lin:2004en} and the Equation of State are input with the lattice simulation results, incorporating with a critical point~\cite{Parotto:2018pwx}. To roughly fit the experimental measurement of Au+Au collisions at 19.6GeV, such as the multiplicity, spectra, and flow, the parameters for the hydrodynamic model are set as~\cite{Tang:2023zvj}: the starting proper time for the AMPT initial condition $\tau_{\tiny 0I}=0.4$fm, the starting time for the hydrodynamic evolution $\tau_{\tiny 0h}=1.2$fm, the specific shear and bulk viscosity $\eta/s=0.08,\zeta/s=0.01$, and the switching energy density from QGP to hadronic phase $e_{\tiny sw}=0.1\mbox{GeV/fm}^3$. The position of the critical point is chosen as $(T_c,\mu_c)=(0.16\mbox{GeV},0.155\mbox{GeV})$. To illustrate the effects of the realistic QGP profile, the simulation is performed with two scenarios:
\begin{enumerate}
    \item \texttt{Scenario I:} The temperature and chemical potential along the longitudinal direction is extracted by averaging the temperature and chemical potential profiles over the whole QGP fireball with the energy density as the weight: $T(\tau,\eta)=\langle T(\tau,\bm{x})\rangle$ and $\mu(\tau,\eta)=\langle \mu(\tau,\bm{x})\rangle$. $T(\tau,\eta)$ and $\mu(\tau,\eta)$ profile from hydrodynamic simulation at 19.6GeV with $0-5\%$ centrality is presented in Figs.~\ref{fig:Tmu}
    \item  \texttt{Scenario II:} For comparison, this work also studies the net-baryon density dynamics within the uniformed background of temperature and chemical potential along the longitudinal direction.  The temperature and chemical potential along the longitudinal direction are fixed the ones at $\eta=0$: $T(\tau,\eta)=\langle T(\tau,x_\perp,\eta=0)\rangle$ and $\mu(\tau,\eta)=\langle \mu(\tau,x_\perp,\eta=0)\rangle$.
\end{enumerate}
As in Ref.~\cite{Tang:2023zvj}, the temperature and chemical potential along the longitudinal direction are obtained by averaging the transverse plane with the energy as weight not simply averaging over the space. This is because the physical quantities ({\it e.g.,} temperature) reflect the macroscopic property of a great number of microscopic particles involved. The appropriate method to obtain the temperature is averaging over the number of particles involved and the energy density is regarded as the quantity approximately proportional to the number of particles. As depicted in the upper panel of Figs. \ref{fig:Tmu}, the temperature $T(\tau,\eta)$ is highest at the center of the fireball and decreases with increasing rapidity $\eta$, and this behavior becomes smoother as a function of proper time $\tau$. Conversely, the chemical potential $\mu_(\tau,\eta)$ reaches its minimum at the fireball's center and exhibits a larger value for higher rapidity. This behavior reflects the baryon-stopping effects along the longitudinal direction. This study only focuses on the dynamics with the QGP fireball, there is no value outside the fireball as the system has freeze-out and turned into the hadronic phase.

\begin{figure}[htb]
  \includegraphics[width=0.48\textwidth]{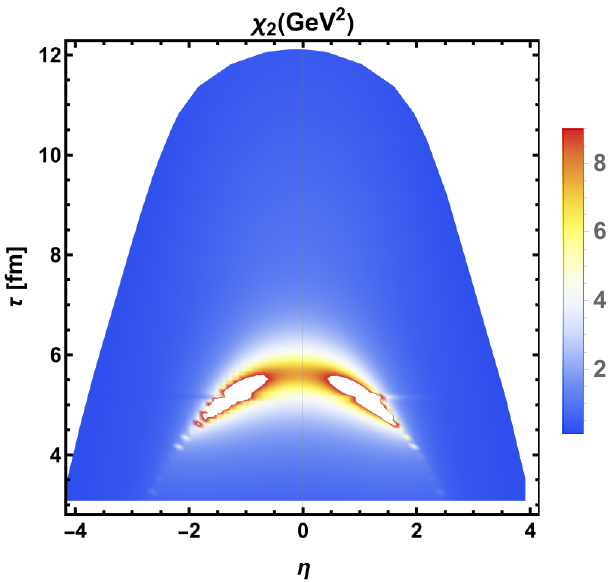}
  \includegraphics[width=0.48\textwidth]{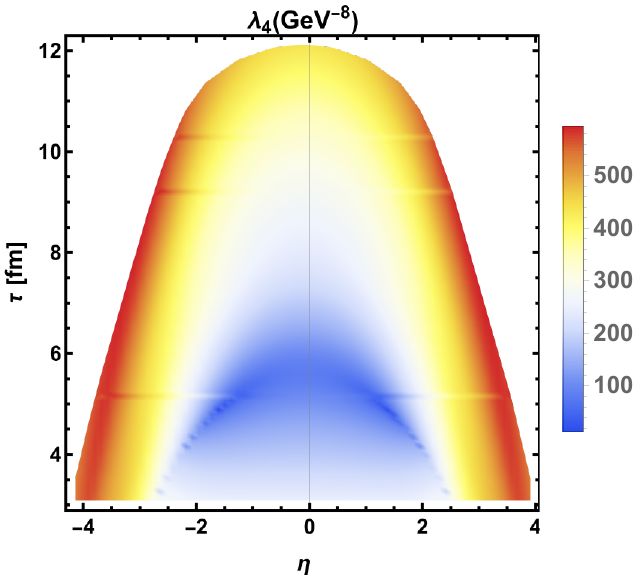}
  \caption{\texttt{Scenario I:}Time evolution of the second order net-baryon susceptibility $\chi_2$ (upper panel) and fourth order coupling coefficient $\lambda_4$ (lower panel) profiles in the rapidity $\eta$ space. Note that $\chi_2(\mbox{GeV}^2)$ in the upper panel only shows a range from 0.1 to 9 GeV$^2$ for illustrative purposes, and the white elliptical regions inside the red region correspond to $\chi_2>9\mbox{GeV}^2$. }
  \label{fig:chi2lambda4QGP}
\end{figure}

\begin{figure}[htb]
  \includegraphics[width=0.48\textwidth]{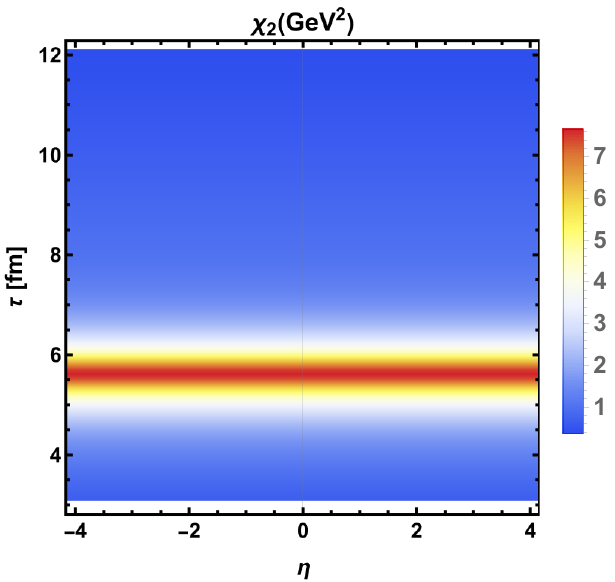}
  \includegraphics[width=0.48\textwidth]{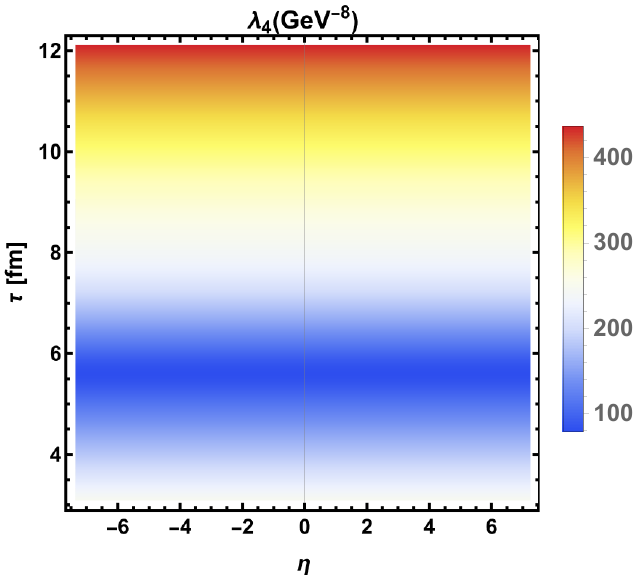}
  \caption{\texttt{Scenario II:} Similar with Fig.\ref{fig:chi2lambda4QGP}, but within the  uniform temperature $T$ and chemical potential $\mu_B$ profile across rapidity $\eta$.}
  \label{fig:chi2lambda4Fix}
\end{figure}

Figs.\ref{fig:chi2lambda4QGP} plot the time evolution of second-order net-baryon susceptibility $\chi_2(\tau,\eta)$ and fourth-order coupling coefficient $\lambda_4(\tau,\eta)$ across the QGP profile. Due to the inhomogeneous temperature and chemical potential across rapidity space~\ref{fig:Tmu}, the local regions at different rapidities reach the critical point $(T_c,\mu_c)$ at varying times. Consequently, the strength of critical effects varies along the rapidity axis for a given proper time. As shown in Fig.\ref{fig:chi2lambda4QGP}, the local regions at finite rapidities (around $\eta=\pm 2$) are the points closest to the QCD critical point, owing to the combined influence of the inhomogeneous $T(\eta)$ and $\mu_B(\eta)$. Therefore, it is natural to expect that multiplicity fluctuations would exhibit nontrivial behavior in response. For the comparison, \texttt{Scenario II}, the time evolution of the coefficients for the uniform temperature and chemical potential has also been shown in Fig.\ref{fig:chi2lambda4Fix}, where the proper time evolution of $\chi_2$ and $\lambda_4$ are uniform across the $\eta$ space.

It requires the solution of the Eq.\eqref{eq:diffeq} to obtain the time evolution of the net-baryon density along the longitudinal axis. This can only be achieved through the numerical simulation of the non-linear equation Eq.\eqref{eq:diffeq} when considering cubic and quartic terms, as well as the inhomogeneous dependence of coefficients ({\it e.g.,} $\chi_2(\tau,\eta)$ and $\lambda_4(\tau,\eta)$) on $\eta$. One of the most popular algorithms for this diffusive stochastic equation is the explicit Forward or Backward Euler Method. However, the Euler Method is conditional stable and the numerical simulation is extremely inefficient near the critical point. This work performs the numerical simulation of Eq.\eqref{eq:diffeq} with the Saul'yev scheme, which is an explicit unconditional stable scheme. In 1957, V.K.Saul’yev proposed the so-called asymmetric methods in the simulation of diffusion equation \cite{Saul'yve:1957,Saul'yve:1964} and has been successfully implemented in the study of Cahn-Hilliard equation~\cite{yang2022explicit}. In addition, the numerical simulation of the diffusion equation \eqref{eq:diffeq} requires the initial configuration of $n_B(\tau_0,\eta)$. In this work, the fluctuations of the baryon density $n_B$ at the beginning of the dynamics are assumed to be in a state of equilibrium, and the initial condition is set as 
\begin{align}
    \langle n_B(\tau_0,\eta)\rangle&=0,\nonumber\\
    \,\langle n_B(\tau_0,\eta)n_B(\tau_0,\eta')\rangle&=\chi_\eta(\tau_0)\delta(\eta-\eta'),
\end{align}
where $\chi_\eta=\chi_2\tau T$. For the detailed numerical implementation of Eq.\eqref{eq:diffeq}, please see Appendix.\ref{app:Numerical}.

Note that the diffusion equation of net-baryon density Eq.~\eqref{eq:diffeq} is obtained with the assumption of the Bjorken flow (boost-invariant) background, while the profiles of the temperature and chemical potential are obtained with the hydrodynamic simulation for the Au+Au collisions at 19.6GeV, which breaks boost-invariance. Such fluid background induces a longitudinal acceleration and results in advection terms on top of the diffusion equation. For theoretical consistency, the diffusion equation Eq.~\eqref{eq:diffeq} requires to be extended to the one with non-boost-invariant fluid. This can be achieved if higher order terms in the effective potential (terms with coefficients $\lambda_3$ and $\lambda_4$) are neglected, but is difficult if those terms are included. Therefore, this work still employs the diffusion equation Eq.~\eqref{eq:diffeq} without considering the background fluid advection effects. Appendix \ref{app:non-boost-invariant} estimates such advection effects, which are found to be moderate for the baryon fluctuations up to the second order.

\section{RESULT AND DISCUSSION}\label{sec:results}
\subsection{Net-baryon number fluctuations within QGP profile}


After solving the dynamics of the conserved net-baryon density \eqref{eq:diffeq}, the configuration of the net-baryon density $n_B(\tau,\eta)$ within the QGP profile is obtained. Therefore, the multiplicity fluctuations within the QGP profile can be calculated as follows:
\begin{align}\label{Eq:cumulants}
    C_1 &= \langle N_B\rangle,\quad    C_2 = \langle (\delta N_B)^2\rangle,\nonumber\\
    C_4 &= \langle (\delta N_B)^4\rangle- 3\langle (\delta N_B)^2\rangle^2,
\end{align}
and 
\begin{align}\label{Eq:cumulants_ratios}
\sigma^2=C_2,\quad\kappa\sigma^2=C_4/C_2.
\end{align}
Here the multiplicity of net-baryon within rapidity interval $\Delta \eta$ at center of rapidity $\bar{\eta}$ is obtained as $N_B=\int^{\bar{\eta}+\Delta \eta}_{\bar{\eta}-\Delta \eta}d\eta n_B(\tau,\eta)$. $\langle \cdots\rangle$ represents averaging over the events and the multiplicity event-by-event fluctuations are defined as $\delta N_B=N_B-\langle N_B\rangle$.

\begin{figure}[htb]
   \includegraphics[width=0.48\textwidth]{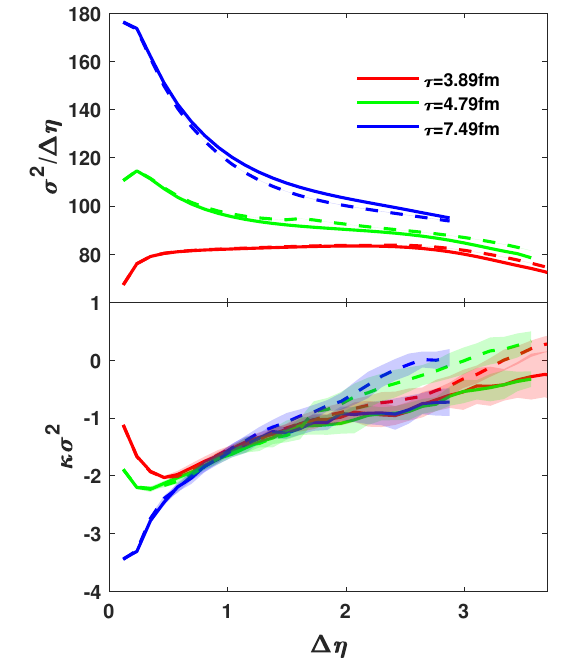}
  \caption{Rapidity acceptance $\Delta \eta$ dependence of the second order multiplicity fluctuations $\sigma^2/\Delta\eta$ (upper panel) and kurotsis $\kappa\sigma^2$ (lower panel) within QGP profile. Curves with different colors correspond to the multiplicity fluctuations at different proper times $\tau$. Dashed for the inhomogeneous profile (\texttt{Scenario I}) and solid for the uniform ones (\texttt{Scenario II}). }
  \label{fig:C2C4Eta}
\end{figure}

\begin{figure}[htb]
  \includegraphics[width=0.48\textwidth]{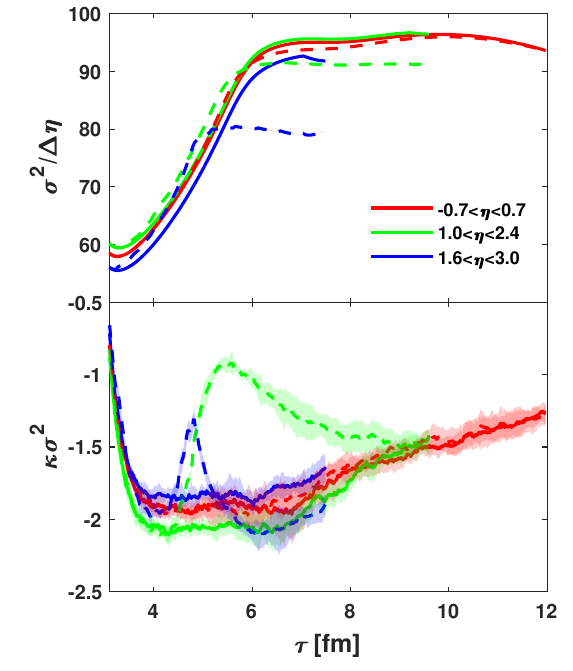}
  \caption{Time evolution of the second order net-baryon susceptibility $\sigma^2/\Delta \eta$ (upper panel) and kurtosis  $\kappa\sigma^2$ (lower panel). Curves with different colors correspond to the multiplicity fluctuations with different rapidity regions $\eta$: $-0.7<\eta<0.7$ ($\bar{\eta}=0$), $1.0<\eta<2.4$ ($\bar{\eta}=1.7$) and $1.6<\eta<3.0$ ($\bar{\eta}=2.3$), with $\Delta\eta=0.7$. Dashed for the inhomogeneous profile (\texttt{Scenario I}) and solid for the uniform ones (\texttt{Scenario II}).}
  \label{fig:C2C4Time}
\end{figure}


Figs.\ref{fig:C2C4Eta} show the rapidity dependence of the net-baryon number fluctuations with fixed proper time $\tau=3.9,4.8$ and $7.5$ fm, respectively. The net-baryon number cumulants $\sigma^2/\Delta\eta$ and $\kappa\sigma^2$ are obtained as in Eq.\eqref{Eq:cumulants_ratios} with the center of rapidity $\bar{\eta}=0$. The second order net-baryon number cumulant $\sigma^2/\Delta \eta$ (kurtosis $\kappa\sigma^2$) shows non-monotonic behavior with increasing rapidity and reaches a maximum (minimum) at small $\Delta\eta$, which agrees with previous studies~\cite{Sakaida:2017rtj,Pihan:2022xcl}. One of the most important properties of the dynamics for the conserved variable is that the diffusion process consumes time (Please see Ref.~\cite{Sakaida:2017rtj} for detail). As a result, the correlation between particles with small $\Delta\eta$ encodes the late-stage dynamics of diffusion, while the one with large $\Delta\eta$ preserves the early evolution of diffusion. As the system scans the critical regime, the susceptibility exhibits a peak with increasing proper time $\tau$. Therefore, the correlation behaves non-monotonically with increasing rapidity interval $\Delta\eta$, which can be regarded as the imprints from the critical fluctuations. Figs.\ref{fig:C2C4Eta} also compares the $\sigma^2/\Delta \eta$ and $\kappa\sigma^2$ in uniform (\texttt{Scenario II}, solid curves) with inhomogeneous (\texttt{Scenario I}, dashed curves) profile. In the case of second-order cumulant $\sigma^2/\Delta\eta$, the difference between \texttt{Scenario II} and \texttt{I} is negligible. For kurtosis, the discrepancy is also small at a small rapidity but becomes significant for large rapidity. This can be understood that the difference of the transport coefficients ($\chi_2$ and $\lambda_4$) in Eq.\eqref{eq:diffeq} between \texttt{Scenario I} and \texttt{II} becomes pronounced at large $\eta$, by comparing Figs.\ref{fig:chi2lambda4QGP} and \ref{fig:chi2lambda4Fix}.

Since the realistic inhomogeneous QGP profile has an impact on the number fluctuations at large rapidity, let's focus on the time evolution of the system fluctuations at different $\eta$ regions. With the number fluctuations obtained from Eq.\eqref{Eq:cumulants}, Figs.\ref{fig:C2C4Time} present the time evolution of $\sigma^2/\Delta\eta$ and $\kappa\sigma^2$ with different $\eta$ intervals both for uniform (\texttt{Scenario II}) and inhomogeneous (\texttt{Scenario I}) profile. Different $\eta$ intervals include $-0.7<\eta<0.7$($\bar{\eta}=0$), $1.0<\eta<2.4$($\bar{\eta}=1.7$) and $1.6<\eta<3.0$($\bar{\eta}=2.3$), with $\Delta\eta=0.7$. In these figures, dashed curves represent the time evolution of the fluctuations with realistic inhomogeneous QGP profile (\texttt{Scenario I}), while the solid curves correspond to the uniform profile case (\texttt{Scenario II}). As shown in Figs.\ref{fig:C2C4Time}, the discrepancy between these two cases is negligible for midrapidity interval ($-0.7<\eta<0.7$), while becomes significant at large rapidity intervals ($1.0<\eta<2.4$ and $1.6<\eta<3.0$)~\footnote{Noted that the integral interval of $\sigma^2$ in Figs.\ref{fig:C2C4Eta} is from $-\Delta\eta$ to $\Delta\eta$. The difference between \texttt{Scenario I} and \texttt{II} is small when the integration is also performed from $-\Delta\eta$ to $\Delta\eta$ for the time evolution of $\sigma^2/\Delta\eta$ in Figs.\ref{fig:C2C4Time}. This is consistent with Figs.\ref{fig:C2C4Eta} and the difference between \texttt{Scenario I} and \texttt{II} in different rapidity is smeared out from $-\Delta\eta$ to $\Delta\eta$. }.

\begin{figure}[htb]
  \includegraphics[width=0.48\textwidth]{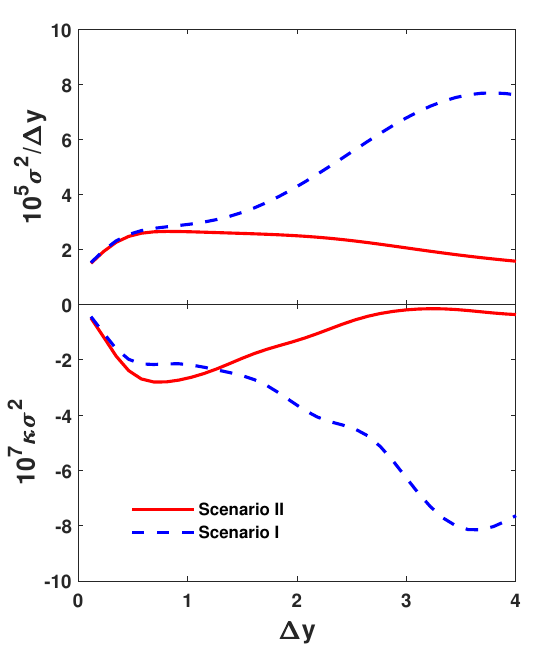}
  \caption{The second order multiplicity fluctuations $\sigma^2/\Delta y$ (upper panel) and kurtosis $\kappa\sigma^2$ (lower panel) along the freeze-out surface as shown in Fig.\ref{fig:Tmu}. Dashed for the inhomogeneous profile (\texttt{Scenario I}) and solid for the uniform ones (\texttt{Scenario II}). }
  \label{fig:C2C4FreezeOut}
\end{figure}

\subsection{Net-baryon number fluctuations at the freeze-out surface}

After the evolution of the net-baryon density with the QGP profile, systems freeze-out and turn into the hadronic phase when the energy density is below the switching energy density $e_{sw}$. The freeze-out hypersurface from the hydrodynamic simulation is employed and shown as the edge of the QGP profile in Figs.\ref{fig:Tmu}. The Cooper-Frye formula~\cite{Cooper:1974mv} is employed to calculate the multiplicity fluctuations after freeze-out:
\begin{align}
    N_B=g\int \frac{d^3p}{(2\pi)^3} \frac{1}{p^0} \int d\sigma_\mu p^\mu f(\bm{x},\bm{p}),
\end{align}
where $g$ is the spin degeneracy and the Boltzmann approximation is taken for the distribution:
\begin{align}
f(\bm{x},\bm{p})=\exp[-(p_\mu u^\mu-\mu)/T].    
\end{align}
As shown in Figs.\ref{fig:Tmu}, the temperature is inhomogeneous across the $\eta$ space and thus fluid cells at different $\eta$ freeze-out with different proper time $\tau_f(\eta)$. This is characterized as the edge of the colored profile. Therefore, the integration measure at constant switching energy density $e_{sw}$ freeze-out hypersurface $\sigma_\mu(x)$ should take the non-boost-invariant form $p^\mu d\sigma_\mu=d\bm{x}_\perp d\eta m_\perp \partial[-\tau_f\sinh(y-\eta)]/\partial\eta$ and $p_\mu u^\mu=m_\perp [u^\tau\cosh(y-\eta)-\tau u^\eta\sinh(y-\eta)]$~\cite{Chattopadhyay:2018dth}. Here $m_\perp =\sqrt{\bm{p}_\perp^2+m^2}$. $\bm{p}_\perp$ and $m$ are transverse momentum and mass of proton, respectively. As pointed out in Sec.\ref{sec:model} and Fig.\ref{fig:utau_ueta}, the term with $\tau u^\eta\sinh(y-\eta)$ is relatively small and is neglected in the following calculation: $p_\mu u^\mu \approx m_\perp \cosh(y-\eta)$ (and $u^\tau \approx 1$). The rapidity distribution of the particle then reduces to:
\begin{align}
 \frac{dN_B}{dy}=\frac{gA}{(2\pi)^2}\int d\eta S(y,\eta) e^{\mu/T} \frac{1}{c^3}\Gamma(3,cm)
\end{align}
where $S(y,\eta)\equiv \partial[-\tau_f \sinh(y-\eta)]/\partial\eta$ and the Gamma function is obtained from $\int dp_\perp p_\perp m_\perp e^{-cm_\perp}=\frac{1}{c^3}e^{-cm}[2+2cm+(cm)^2]\equiv \frac{1}{c^3}\Gamma(3,cm)$ with $c\equiv \cosh(y-\eta)/T$. The integral over the transverse space $dx_\perp$ is represented by $A=1\mbox{fm}^2$~\cite{Nahrgang:2020yxm}. Replacing $\mu$ by $-\mu$ in the above equations gives the anti-baryon distribution $dN_{\bar{B}}/dy$. As there is only one fluctuating variable $n_B$ in this framework, following Ref.\cite{Ling:2013ksb}, the multiplicity fluctuations is obtained as
\begin{align}\label{eq:fluctAtFreezeOut}
    \delta \biggl(\frac{dN_{B-\bar{B}}}{dy}\biggr)&=\frac{gA}{(2\pi)^2}\int d\eta S(y,\eta) \frac{\delta n_B}{T\tau_f\chi_2} \frac{1}{c^3}\Gamma(3,cm)\nonumber\\
    &\qquad\qquad\times [\exp(\mu/T)+\exp(-\mu/T)],
\end{align}
by considering the definition of the susceptibility $\delta \mu =\delta n_B(\tau_f,\eta)/(\tau_f \chi_2)$ in Milne frame. Here, the integration over $\eta$ is performed at the edge of the hydrodynamic profile (the freeze-out surface). The fluctuations of the net-baryon density $n_B(\tau_f,\eta)$, as well as the temperature $T(\tau_f,\eta)$, chemical potential $\mu(\tau_f,\eta)$ and susceptibility $\chi_2(\tau_f,\eta)$ are also extracted from the hypersurface. The fluctuations of net-baryon are obtained by integration over $y$: $\delta N_{B-\bar{B}} = \int^{\Delta y}_{-\Delta y}dy\delta(dN_{B-\bar{B}}/dy)$ and various orders of the cumulants of the multiplicity fluctuations are calculated as Eq.\eqref{Eq:cumulants} accordingly.


Following Eq.\eqref{eq:fluctAtFreezeOut}, net-baryon multiplicity fluctuations at the freeze-out surface are calculated and shown in Figs.\ref{fig:C2C4FreezeOut}.   For illustrative purposes, Figs.\ref{fig:C2C4FreezeOut} also presents the multiplicity fluctuations obtained with the uniform $T$ and $\mu$ profile (\texttt{Scenario II}) but still at the edge of the profile \ref{fig:Tmu}. Note that the freeze-out surface for \texttt{Scenario II} is not the edge of the profile \ref{fig:Tmu}, and the solid curves are presented for comparison. In the scenario of a uniform temperature and chemical potential profile, second order fluctuations $\sigma^2/\Delta y$ (kurtosis $\kappa\sigma^2$) display an increase (decrease) first followed by a rapid decline (increase) with increasing rapidity interval $\Delta y$, exhibiting a maximum (minimum) at small values of rapidity interval $\Delta y$. This behavior is consistent with studies on diffusive dynamics near the critical point~\cite{Sakaida:2017rtj,Pihan:2022xcl}, suggesting that the correlation with large $\Delta y$ preserves the early dynamics of diffusion. On the contrary, $\sigma^2/\Delta y
$ and $\kappa\sigma^2$ in the case of the realistic QGP profile deviate significantly with uniform temperature and chemical potential case at large rapidity interval $\Delta y$. As shown in Figs.\ref{fig:Tmu}, the fluctuations with different rapidity at the freeze-out surface are determined by the dynamics with different time $\tau_f(\eta)$. The significant enhancement (decrease) of $\sigma^2/\Delta y$ and $\kappa\sigma^2$ is a result of the dynamics occurring at small $\tau_f$.

\begin{figure}[htb]
  \includegraphics[width=0.5\textwidth]{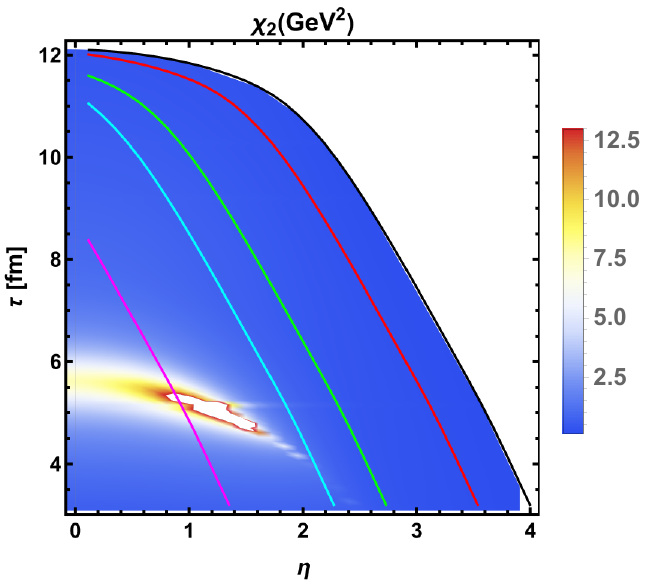}
  \caption{A sketch of the curves before freeze-out. Along these curves, the susceptibility $\chi_2$ and second order density cumulants $C_2/\Delta \eta$ are extracted (See Fig.\ref{fig:chi2C2curve}) to study the diffusion of the fluctuations before freeze-out.}
  \label{fig:curves}
\end{figure}

\begin{figure}[htb]
  \includegraphics[width=0.48\textwidth]{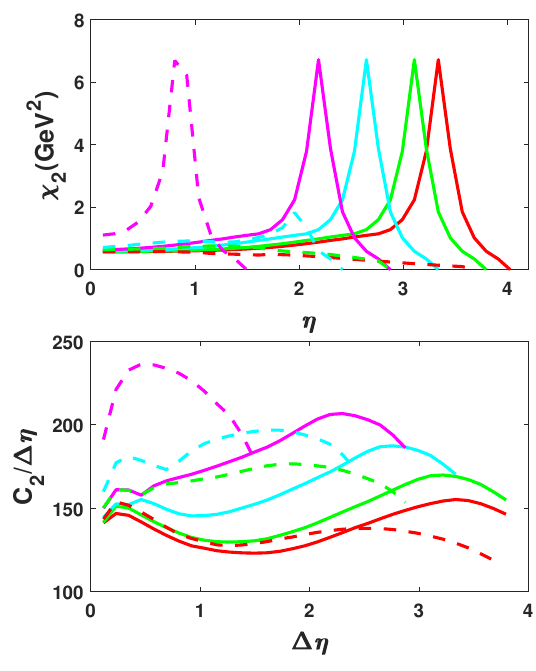}
  \caption{Dashed: The susceptibility $\chi_2$ (upper panel) and second order density cumulants $C_2/\Delta \eta$ (lower panel) along the curves in Fig.\ref{fig:curves}. $\chi_2$ and $C_2/\Delta\eta$ with different colors are extracted along the corresponding curves in Fig.\ref{fig:curves}(\texttt{Scenario I}). Solid: The susceptibility $\chi_2$ (upper panel) and second order density cumulants $C_2/\Delta \eta$ (lower panel) along the curves in Fig.\ref{fig:curves} with $\chi_2$ profile replaced by the Fig.\ref{fig:chi2lambda4Fix} (\texttt{Scenario II}).}
  \label{fig:chi2C2curve}
\end{figure}

To see this, it is instructive to study the rapidity dependence of the multiplicity fluctuations before freeze-out. This can be achieved by moving the freeze surface backward to the smaller $(\tau,\eta)$ region. In Fig.\ref{fig:curves}, the curves in QGP fireball are depicted, along which the variables at smaller $(\tau,\eta)$ region are extracted. Figs.\ref{fig:chi2C2curve} show the susceptibility $\chi_2$ and second-order cumulant $C_2/\Delta\eta$ along the curves in Fig.\ref{fig:curves}. Here $C_2=\langle [\int^{\Delta \eta}_{-\Delta\eta}d\eta \delta n_B]^2\rangle$, where the integral is performed along the curves in Fig.\ref{fig:curves}. With increasing $\tau$ in \texttt{Scenario I}, one could see the peak of susceptibility $\chi_2$ only exhibits along the curves with small $\tau$ and $\eta$, and been significantly suppressed near the freeze-out surface. This is the impact of the inhomogeneous QGP profile, in which the system passes through the critical regime and a pulse of the susceptibility appears and vanishes rapidly. On the other hand, during the process of diffusion, the pulse of the susceptibility near the QCD critical point results in the large fluctuations $C_2$ at small $\Delta\eta$, as shown in the dashed-magenta curve in the lower panel of Figs.\ref{fig:chi2C2curve}. The large fluctuations $C_2$ smears and diffuses with increasing proper time and rapidity, but can not catch up with the rapid decreasing $\chi_2$, which is known as the critical slowing down effects. In short, the evolution of the $T$ and $\mu$ profiles drives the susceptibility passing through the critical point, while the critical slowing down effects preserve the memory of the large net-baryon density fluctuations at the critical point. Consequently, the fluctuations in Eq.\eqref{eq:fluctAtFreezeOut} behave with pronounced enhancement at large rapidity intervals. On the contrary, one could see the susceptibility $\chi_2$ always has a peak and moves to larger $\eta$ in \texttt{Scenario II}, which is expected in Fig.\ref{fig:chi2lambda4Fix}. Therefore, the fluctuations at Eq.\eqref{eq:fluctAtFreezeOut} drop down rapidly with increasing $\Delta \eta$ even with the diffusion of net-baryon density (lower panel of Figs.\ref{fig:chi2C2curve}).

\section{Conclusion and Outlook}\label{conclusion}
This work studies the conserved dynamics of the net-baryon multiplicity fluctuations near the QCD critical point with the inhomogeneous temperature and chemical potential profile. The dynamics of conserved net-baryon density are based on the numerical simulation of the 1+1-dimensional stochastic diffusion equation. The realistic QGP profile is obtained by the hydrodynamic simulation and the susceptibilities in the diffusion equation are constructed based on the temperature and chemical potential profile. In this context, the QGP profile is inhomogeneous and the susceptibilities are non-trivial across the proper time $\tau$ and rapidity $\eta$ plane. As discovered by the early studies~\cite{Sakaida:2017rtj}, the fluctuations of the conserved net-baryon density behave non-monotonically with the increasing rapidity interval, because the correlation function in the diffusion process preserves the evolution history at a large rapidity interval. It is found that the influence of the inhomogeneous profile is negligible at small rapidity but relatively pronounced at large rapidity, as the susceptibility in the realistic case deviates from the case of the uniform profile at large rapidity. Furthermore, the fluctuations on the freeze-out hypersurface have also been investigated. Comparing the case with a uniform QGP profile, the magnitudes of second-order cumulants as well as kurtosis in inhomogeneous profile present significant enhancement at large rapidity. This is the result of the combined effect of critical slowing down and inhomogeneous profile.

Finally, it deserves to be pointed out that this study of the critical fluctuations is based on a simplified model, where only the 1+1-dimensional conserved net-baryon density is considered. Other degrees of freedom in hydrodynamic evolution have not been considered comprehensively, and are only regarded as the background. As shown in this paper, the dynamical simulation of the relevant quantity with a realistic setup is essential for comparison with experimental measurement. For example, the hydrodynamics coupling with the additional slow modes has been developed~\cite{Stephanov:2017ghc}, and further incorporating the higher order slow modes, their extension at freeze-out surface~\cite{Pradeep:2022eil}, and corresponding phenomenological study are required for the persuasive prediction. Recently, the analytical structure of the higher-order slow modes has also been discussed~\cite{2024arXiv241007929A}. In this work, the dynamical model for the conserved baryon density is constructed based on the boost-invariant fluid background, which is relatively acceptable for the intermediate collision energy of the Beam Energy Scan program. The dynamical model is required to be extended to the non-boost-invariant context for the system approaching the lower energies.

\section*{Acknowledgments}
The author would like to thank Shian Tang for his contribution at the beginning of this study and discussion with Huichao Song, Navid Abbasi and Masakiyo Kitazawa. This work is supported by the NSFC under grant No. 12305143 and the China Postdoctoral Science Foundation under Grant No. 2023M731467.

\appendix

\section{Parametrization}\label{app:parametrization}
The coefficients of the Eq.\eqref{eq:diffeq} require to be specified for the simulation of net-baryon density near the QCD critical point. According to the universal analysis, one can obtain the diffusion coefficient $D$, second-order baryon susceptibility $\chi_2$, and third- and fourth-order coupling coefficients $\lambda_3,\lambda_4$ from the mapping three-dimensional Ising model. Following Refs.~\cite{Sakaida:2017rtj,Pihan:2022xcl}, these coefficients include regular and singular parts:
\begin{align}\label{eq:coefficients}
    \chi_2&=\chi^{\mbox{cri}}_2+\chi^{\mbox{reg}}_2,\nonumber\\
    \lambda_3&=\lambda^{\mbox{cri}}_3+\lambda^{\mbox{reg}}_3,\nonumber\\
    \lambda_4&=\lambda^{\mbox{cri}}_4+\lambda^{\mbox{reg}}_4,
\end{align}
where the regular coefficients read
\begin{align}
    \lambda^{\mbox{reg}}_3=-\frac{1}{2}\chi^{\mbox{reg}}_3\chi^{-3}_2,\quad \lambda^{\mbox{reg}}_4=\frac{1}{2} (\chi^{\mbox{reg}}_3)^2\chi^{-5}_2-\frac{1}{6} \chi^{\mbox{reg}}_4\chi^{-4}_2.\nonumber
\end{align}
With the susceptibility, one has the diffusion coefficient $D=D_c/\chi_2$ near the critical point. $D_c$ and surface tension coefficient are treated as constants: $D_c=16.5$ and $K=0.005$.

The regular susceptibilities are obtained by interpolating interpolate between the regular susceptibilities at hadronic phase $\chi^{\mbox{\tiny H}}_n$ and QGP phase $\chi^{\mbox{\tiny QGP}}_n$
\begin{align}
    \chi^{\mbox{reg}}_n=\chi^{\mbox{\tiny H}}_n+(\chi^{\mbox{\tiny QGP}}_n-\chi^{\mbox{\tiny H}}_n)S(T),\nonumber
\end{align}
where the interpolating function is $S(T)=\frac{1}{2}(1+\tanh((T-T_c)/\delta T))$ and the width of the transition region is set as $\delta T=0.01$GeV. Following Ref.~\cite{Ling:2013ksb}, the second-order susceptibility $\chi^{\mbox{\tiny H,latt}}_2/T^2_{\mbox{\tiny H}}=1/3$ for hadronic phase and $\chi^{\mbox{\tiny QGP,latt}}_2/T^2_{\mbox{\tiny QGP}}=2/3$ for QGP phase, the low-temperature limit is set as $T_{\mbox{\tiny H}}=0.1$GeV and the high-temperature limit $T_{\mbox{\tiny QGP}}=0.25$GeV. For the fourth order susceptibility, $\chi^{\mbox{\tiny H,latt}}_4/\chi^{\mbox{\tiny H,latt}}_2=1$ at hadronic phase and $\chi^{\mbox{\tiny QGP,latt}}_4/\chi^{\mbox{\tiny QGP,latt}}_2=2/(3\pi^2)$ from the lattice simulation ~\cite{Cheng:2008zh,Bazavov:2017dus}. The third-order susceptibility is assumed with $\chi^{\mbox{\tiny latt}}_3/\chi^{\mbox{\tiny latt}}_2=\chi^{\mbox{\tiny latt}}_4/\chi^{\mbox{\tiny latt}}_2$ from Refs.~\cite{Motornenko:2019arp,Mukherjee:2016nhb}. Finally, one has the regular susceptibilities both for the hadronic and QGP phases:
\begin{align}
    \chi^{\mbox{\tiny H}}_2&=T^2_{\mbox{\tiny H}} \chi^{\mbox{\tiny H,latt}}_2,\quad \chi^{\mbox{\tiny QGP}}_2=T^2_{\mbox{\tiny QGP}} \chi^{\mbox{\tiny QGP,latt}}_2,\nonumber\\
    \chi^{\mbox{\tiny H}}_3&=T_{\mbox{\tiny H}} \chi^{\mbox{\tiny H,latt}}_3,\quad \chi^{\mbox{\tiny QGP}}_3=T_{\mbox{\tiny QGP}} \chi^{\mbox{\tiny QGP,latt}}_3,\nonumber\\
    \chi^{\mbox{\tiny H}}_4&= \chi^{\mbox{\tiny H,latt}}_4,\qquad \chi^{\mbox{\tiny QGP}}_4= \chi^{\mbox{\tiny QGP,latt}}_4.\nonumber
\end{align}

The critical contributions $\chi^{\mbox{\tiny cri}}_2,\lambda^{\mbox{\tiny cri}}_3$ and $\lambda_4^{\mbox{\tiny cri}}$ are constructed from the cumulants of Ising model $\kappa_2$
\begin{align}
    \chi^{\mbox{\tiny cri}}_2&=T^2_AC_c\kappa_2\equiv\xi^2, \quad \lambda^{\mbox{\tiny cri}}_3=\tilde{\lambda}_3T(T\xi)^{-3/2}/T^3_A,\nonumber\\
    \lambda_4^{\mbox{\tiny cri}} &= \tilde{\lambda}_4(T\xi)^{-1}/T^6_A,
\end{align}
where $C_c$ is mapping constant and chosen as $C_c=4$ in this work. $T_A$ is the constant for the dimensional consistency and is set as $T_A=0.5$GeV. The dimensionaless coupling constants $\tilde{\lambda}_3$ and $\tilde{\lambda}_4$ have values range (0,8) and (4,20), respectively~\cite{Tsypin:1994nh}. $\tilde{\lambda}_3=1$ and $\tilde{\lambda}_4=4$ are used in this study.  The second-order cumulant of the Ising model reads~\cite{Nonaka:2004pg,Sakaida:2017rtj}
\begin{align}
    \kappa_2=\frac{M_0}{H_0} \frac{1}{R^{4/3}(3+2\theta^2)},
\end{align}
where the normalization constants are $M_0\simeq0.605,h_0\simeq 0.394$. On the phase diagram near the critical point, the distance $R$ and angle $\theta$ to the critical point are calculated with the equations
\begin{align}
    r(R,\theta)=R(1-\theta^2),\quad h(R,\theta) =R^{5/2}(3\theta-2\theta^3).
\end{align}
The Ising variables ($r,h$) are connected to the temperature and chemical potential of the QCD system by the mapping:
\begin{align}
    \frac{T-T_c}{\Delta T}=\frac{h}{\Delta h},\quad \frac{\mu-\mu_c}{\Delta \mu}=-\frac{r}{\Delta r},
\end{align}
where $T_c$ and $\mu_c$ are the critical temperature and chemical potential, respectively. $\Delta T$ and $\Delta \mu$ represent the widths of the critical region of the QCD phase diagram, $\Delta h$ and $\Delta r$ are the ones in the Ising model. These are non-universal parameters and are set as $\Delta T=T_c/8,\Delta\mu=0.1$GeV,$\Delta r=(5/3)^{3/4},\Delta h=1$ in this work. The temperature $T(\tau,\eta)$ and chemical potential $\mu(\tau,\eta)$ of the QGP profile are borrowed from hydrodynamic simulation, which is addressed in Sec.~\ref{sec:model}

\section{Numerical details of Eq.\eqref{eq:diffeq} }\label{app:Numerical}
This appendix presents the details of the numerical simulation of Eq.\eqref{eq:diffeq} and its verification by comparison with the analytical calculation in the linear limit of Eq.\eqref{eq:diffeq}.

The explicit form of Eq.\eqref{eq:diffeq} reads
\begin{align}
    \partial_\tau n_B&=D_\eta \chi_\eta \frac{\partial^2}{\partial \eta^2} \biggl[\frac{n_B}{\chi_\eta}-\frac{K}{\tau^3}\frac{\partial^2}{\partial \eta^2}n_B+\frac{\lambda_3}{\tau^2}n_B^2+\frac{\lambda_4}{\tau^3}n_B^3\biggr]\nonumber\\
    &\qquad-\partial_\eta \zeta(\tau,\eta),
\end{align}
with the noise correlator
\begin{align}\label{eq:noiseAppendix}
    \langle \zeta(\tau,\eta)\zeta(\tau',\eta')\rangle =2D_\eta\chi_\eta\delta(\tau-\tau')\delta(\eta-\eta'),
\end{align}
where $D_\eta=D/\tau^2,\chi_\eta=\chi_2\tau T$. 

\begin{figure}[htb]
  \includegraphics[width=0.48\textwidth]{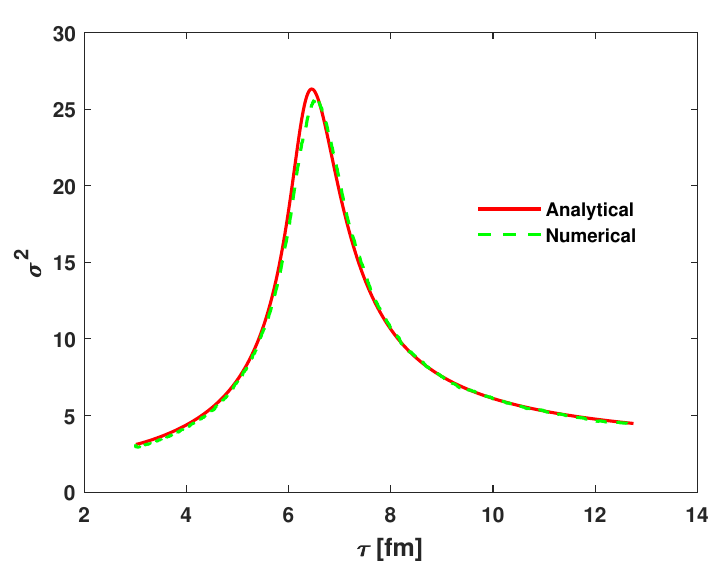}
  \caption{Time evolution of second-order net-baryon multiplicity fluctuations $\sigma^2$ as a solution of the dynamics of the net-baryon density without higher order terms \eqref{eq:linearEq}. The green dashed curve of $\sigma^2$ is obtained from the simulation with the Saul'yev scheme~\eqref{eq:Saulyev} and the red one corresponds to the analytical calculation of Eq.~\eqref{eq:analytical}. $\Delta \eta=0.25$ here.}
  \label{fig:C2analytical}
\end{figure}

One of the most popular algorithms for this diffusive stochastic equation is the explicit Forward or Backward Euler Method. In this numerical algorithm, Eq.\eqref{eq:diffeq} can be discretized in an explicit form: $(n^{i+1}_{B,j}-n^i_{B,j})/\Delta \tau=f(n^i_{B,j})$, where the next time step $n^{i+1}_{B,j}$ only shown in the left-hand side and can be obtained explicitly. However, this scheme is conditionally stable because it is only an approximation to Eq.\eqref{eq:diffeq}, error will gradually accumulate 
and eventually lead to instability. The situation is exacerbated as the system approaches the critical point. To attain stable solutions, the temporal step should be significantly smaller than the spatial step, such as $\Delta \tau\leq \Delta \eta^2/(2D)$ for diffusion equation $\partial \tau n_B=D\partial^2_\eta n_B$, namely the conditional stable. This makes the numerical simulation extremely inefficient.

This work implements the numerical simulation with Saul'yev scheme~\cite{Saul'yve:1957,Saul'yve:1964}, where this equation can be  discretized as 
\begin{align}\label{eq:Saulyev}
    n^{i+1}_j&\approx \frac{1}{r}\biggl\{\biggl(1-D_\eta\frac{\delta\tau}{\delta \eta^2}\biggr)n^i_j+D_\eta\frac{\delta \tau}{\delta \eta^2}\biggl(n^i_{j+1}+n^{i+1}_{j-1}\biggr)\nonumber\\
    &\quad+D_\eta \chi_\eta\frac{\delta \tau}{\delta \eta^2}\biggl(S[n^i_{j+1}]+S[n^i_{j-1}]-2S[n^i_{j}]\biggr)\nonumber\\
    &\quad-D_\eta\chi_\eta\frac{K}{\tau_i^3}\frac{\delta \tau}{\delta \eta^4}\biggl(n^{i}_{j+2}-4n^{i}_{j+1}-4n^{i+1}_{j-1}+n^{i+1}_{j-2}\biggr)\nonumber\\
    &\quad+\sqrt{\frac{2D_\eta\chi_\eta}{A\delta \tau \delta \eta}} \frac{\delta \tau}{\delta \eta} \biggl(W^i_{j+1}-W^i_{j}\biggr)\biggr\},
\end{align}
where $r=(1+D_\eta\frac{\delta \tau}{\delta \eta^2}+6D_\eta \chi_\eta \frac{K}{\tau^3}\frac{\delta\tau}{\delta \eta^4})$ and $S[n^i_j]$ denotes the discretization of the higher order terms $\frac{\lambda_3}{\tau^2}n_B^2+\frac{\lambda_4}{\tau^3}n_B^3$. The noise term is discretized into $W^i_j$, corresponding to the Gaussian white noise with unit variance. $n^i_j$ is the net-baryon density at proper time $\tau_i=\tau_0+i\cdot \delta\tau$ and rapidity $\eta_j=-L/2+j\cdot \delta \eta$. In this simulation, the increment in the proper time is chosen as $\delta\tau=0.0018$fm and the spacing of the rapidity is $\delta \eta=0.115$. The grid size of the simulation is $L=\delta \eta \cdot N$ with $N=128$. The initial proper time is $\tau_0=3$fm with the initial condition: $\langle n_B(\tau_0,\eta)\rangle=0,\,\langle n_B(\tau_0,\eta)n_B(\tau_0,\eta')\rangle=\chi_\eta(\tau_0)\delta(\eta-\eta')$.

As shown in Eq.\eqref{eq:Saulyev}, the net-baryon density for the next temporal step $i+1$ at grid site $j$ only depends on the ones of time step $i$, expect $n^{i+1}_{j-1}$ and $n^{i+1}_{j-2}$ at right hand side of Eq.\eqref{eq:Saulyev}. Therefore, $n^{i+1}_j$ can be computed explicitly, with the boundary condition of the cell $j=1$  given, and the simulation starts the cell from the left to the right $j=1,\cdots, N$. On the other hand, the freeze-out hyper-surface of the QGP profile naturally provides the boundary of the simulation. At the edge of this hypersurface, I employ the boundary condition as follows: evolving the net-baryon density with Eq.\eqref{eq:Saulyev}, but replacing the cell outside (e.g.,$n^i_{j+1}$) with the closest boundary (e.g.,$n^i_{j}$) if the cell $i$ fall at the right side of the profile.

To verify the implemented numerical scheme Eq.\eqref{eq:Saulyev}, the dynamics of the net-baryon density without the higher order terms are investigated:
\begin{align}\label{eq:linearEq}
    \partial_\tau n_B&=D_\eta \frac{\partial^2}{\partial \eta^2} n_B-\partial_\eta \zeta(\tau,\eta),
\end{align}
with noise~\eqref{eq:noiseAppendix}. This is the stochastic diffusion equation employed in the previous works~\cite{Sakaida:2017rtj,Wu:2019qfz} and the corresponding analytical solution of second-order baryon multiplicity fluctuations reads
\begin{align}\label{eq:analytical}
    \sigma^2&=\Delta \eta \chi_\eta(\tau)-\Delta \eta\int^\tau_{\tau_0}d\tau_1\chi'_\eta(\tau_1)\biggl\{\mbox{erf}\biggl[\frac{1}{\bar{D}(\tau_1,\tau,\Delta \eta)}\biggr]\nonumber\\
    &\qquad+\frac{\bar{D}(\tau_1,\tau,\Delta \eta)}{\sqrt{\pi}}\biggl[\exp\biggl(-\frac{1}{\bar{D}(\tau_1,\tau,\Delta \eta)^{2}}\biggr)-1\biggr]\biggr\},
\end{align}
where $\bar{D}(\tau_1,\tau,\Delta \eta)=\Delta\eta/\sqrt{8\int^\tau_{\tau_1}d\tau'D_\eta(\tau')}$ and $\chi'_\eta(\tau_1)$ is the derivative of $\chi_\eta$ over the proper time.

It requires the time evolution of $\chi_\eta$ and $D_\eta$ for the comparison of analytical second order multiplicity Eq.~\eqref{eq:analytical} with the numerical results of Eq.~\eqref{eq:linearEq} with the algorithm Eq.\eqref{eq:Saulyev}. This can be constructed following the method of Eq.~\eqref{eq:coefficients} in Append~\ref{app:parametrization}. Here, the time evolution of temperature $T(\tau,\eta)$ is not obtained from QGP profile but is employed with the Hubble-like expansion~\cite{Sakaida:2017rtj,Wu:2019qfz}, $T(\tau)=T_I(\tau/\tau_0)^{-c^2_s}$ and fixed chemical potential $\mu=0.13$GeV, with $T_I=0.325$GeV and $c^2_s=1$. In this calculation, temperature and chemical potential are constant across the rapidity space. Fig.\ref{fig:C2analytical} shows the comparison of the time evolution of $\sigma^2$ obtained both from numerical simulation and analytical calculation. The agreement between these two methods verifies the reliability of the Saul'yev scheme in the simulation of Eq.~\eqref{eq:diffeq}.

\section{The extension of the diffusion equation to the non-boost-invariant fluid }\label{app:non-boost-invariant}

\begin{figure}[htb]
  \includegraphics[width=0.48\textwidth]{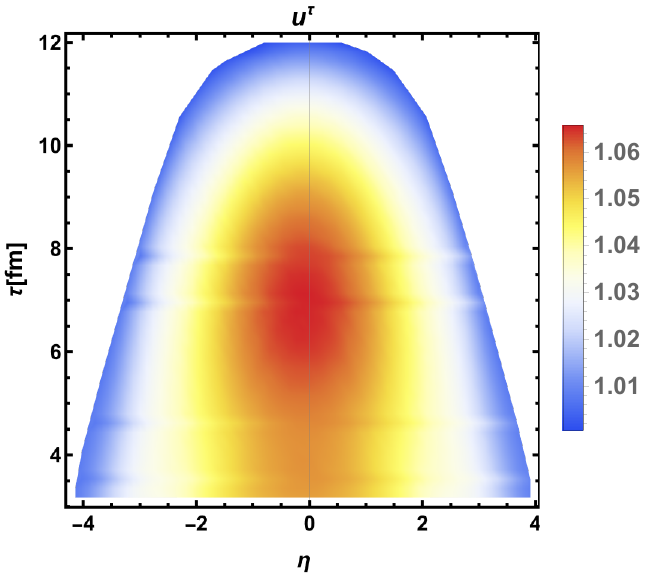}
  \includegraphics[width=0.48\textwidth]{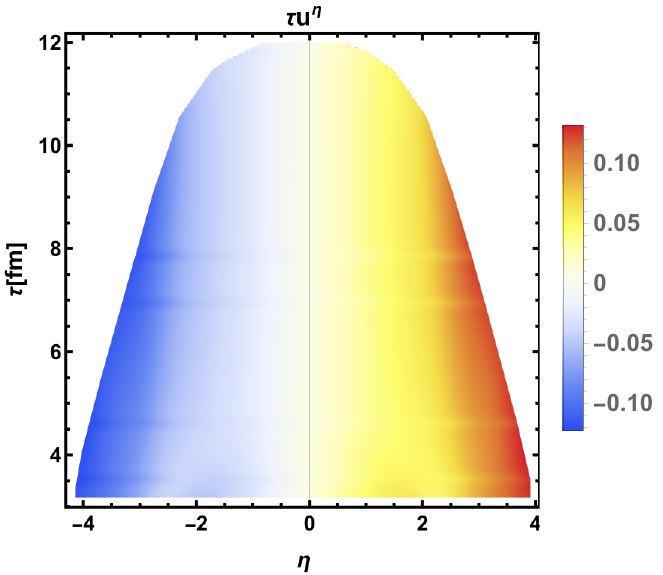}
  \caption{Similar with Fig.\ref{fig:Tmu} but for the flow velocity $u^\tau$ (upper panel) and  $\tau u^\eta$ (lower panel) profiles.}
  \label{fig:utau_ueta}
\end{figure}

As pointed out in Sec.\ref{sec:model}, the stochastic diffusion equation \eqref{eq:diffeq} employed in this work are obtained within the background of the Bjorken flow. This requires to be justified and discussion of the possible extension to a more general context ({\it e.g.,} within the non-boost-invariant background).

In this appendix, the non-boost-invariant fluid background is taken as
\begin{align}\label{Eq:non-boost-invariant_flow}
    u^\mu=(u^\tau\cosh{\eta}+\tau u^\eta \sinh{\eta},0,0,u^\tau\sinh{\eta}+\tau u^\eta \cosh{\eta}),
\end{align}
where the velocity in the transverse plane is neglected ($u^x=0,u^y=0$). Consider the conservation equation for the current
\begin{align}\label{Eq:current_conservation}
    \partial_\mu(n_Bu^\mu) + D\chi_2 T\partial_\mu [\Delta^{\mu\nu}\partial_\nu(\beta \mu_B)]+\partial_\mu \zeta^\mu=0,
\end{align}
within the fluid background \eqref{Eq:non-boost-invariant_flow}, the first and second term are obtained as
\begin{align}
    \partial_\mu (n_Bu^\mu) =u^\tau \frac{\partial n_B}{\partial \tau} + u^\eta \frac{\partial n_B}{\partial \eta} + n_B\bigg[\frac{\partial u^\tau}{\partial \tau}+\frac{u^\tau}{\tau}+\frac{1}{\tau}\frac{\partial (\tau u^\eta)}{\partial \eta}\bigg],
\end{align}
and 
\begin{align}\label{Eq:diffusion_term_non_boost}
    &\partial_\mu [\Delta^{\mu\nu}\partial_\nu(\beta \mu_B)]\nonumber\\
    &=-\frac{\partial(\beta\mu_B)}{\partial \tau}\bigg[2\tau u^\eta \frac{\partial \tau u^\eta}{\partial \tau}+\frac{(\tau u^\eta)^2}{\tau}+\frac{u^\tau}{\tau}\frac{\partial \tau u^\eta}{\partial \eta}+\frac{\tau u^\eta}{\tau} \frac{\partial u^\tau}{\partial \eta}\bigg]\nonumber\\
    &\quad -\frac{\partial (\beta \mu_B)}{\partial\eta}\bigg[\frac{u^\tau}{\tau} \frac{\partial(\tau u^\eta)}{\partial \tau}+2\frac{\tau u^\eta}{\tau^2} \frac{\partial(\tau u^\eta)}{\partial \eta} +\frac{\tau u^\eta}{\tau} \frac{\partial u^\tau}{\partial \tau}\bigg]\nonumber\\
    &\quad -\frac{(u^\tau)^2}{\tau^2} \frac{\partial^2 (\beta \mu_B)}{\partial \eta^2}-(\tau u^\eta)^2 \frac{\partial^2(\beta \mu_B)}{\partial\tau^2}-2\frac{\partial^2(\beta \mu_B)}{\partial\tau\partial\eta} \frac{u^\tau}{\tau} (\tau u^\eta),
\end{align}
where $\beta\equiv1/T$ and $\mu_B=\delta F/\delta n_B$. In this calculation, the last two terms in Eq.\eqref{Eq:diffusion_term_non_boost} are neglected considering $\tau u^\eta$ is small. In principle, the effective potential $F$ near the critical point should take into account all the higher order terms in Eq.\eqref{Eq:effective_potential}. While it is hard to implement in the dynamics with the non-boost-invariant fluid background \eqref{Eq:current_conservation}, and only the second order term in Eq.\eqref{Eq:effective_potential} are considered in this appendix by assuming the system is not extremely close to the critical point:
\begin{align}
    \beta \mu_B \approx n_B/(T\tau \chi_2).
\end{align}
Therefore, the equation of the dynamics of fluctuations within the non-boost-invariant fluid background becomes
\begin{align}\label{Eq:diffusion_non_boost_invariant}
    \frac{\partial n_B}{\partial \tau} = \frac{1}{\alpha_0}\biggl[\bigg(\frac{u^\tau}{\tau}\bigg)^2 \frac{D}{\tau} \frac{\partial^2 n_B}{\partial \eta^2} - \beta_0 \frac{\partial n_B}{\partial \eta} -\gamma_0 n_B- \frac{1}{\tau} \partial_\eta \zeta^\eta\biggr],
\end{align}
with the noise term in Eq.\eqref{Eq:noise} and 
\begin{align}
  \alpha_0 &\equiv \frac{u^\tau}{\tau} - \frac{D}{\tau} \bigg[2(\tau u^\eta) \frac{\partial(\tau u^\eta)}{\partial \tau} +\frac{u^\tau}{\tau }\frac{\partial(\tau u^\eta)}{\partial \eta}+u^\eta \frac{\partial u^\tau}{\partial \eta}\bigg],\nonumber\\
  \beta_0&\equiv \frac{u^\eta}{\tau} -\frac{D}{\tau} \bigg[\frac{u^\tau}{\tau} \frac{\partial(\tau u^\eta)}{\partial \tau}+2\frac{u^\eta}{\tau}\frac{\partial (\tau u^\eta)}{\partial \eta}+\frac{\tau u^\eta}{\tau} \frac{\partial u^\tau}{\partial \tau}\bigg],\nonumber\\
  \gamma_0&\equiv\frac{1}{\tau} \frac{\partial u^\tau}{\partial \tau} + \frac{1}{\tau^2} \frac{\partial (\tau u^\eta)}{\partial \eta}\nonumber\\
  &\qquad+ \frac{D}{\tau} \bigg[2(\tau u^\eta) \frac{\partial (\tau u^\eta)}{\partial \tau}+\frac{u^\tau}{\tau}\frac{\partial(\tau u^\eta)}{\partial \eta}+u^\eta \frac{\partial u^\tau}{\partial \eta}\bigg].\nonumber
\end{align}
The equation \eqref{Eq:diffusion_non_boost_invariant} is the so-called advection-diffusion equation. In the boost-invariant limit: $u^\tau\rightarrow1,\tau u^\eta\rightarrow0$, then $\alpha_0\rightarrow 1/\tau,
\beta_0\rightarrow0,\gamma_0\rightarrow0$, Eq.~\eqref{Eq:diffusion_non_boost_invariant} reproduces the equation without the higher order terms Eq.~\eqref{eq:linearEq}.  The non-boost-invariant fluid acceleration induces the advection effects in the dynamics of net-baryon density, as shown in the additional terms in Eq.~\eqref{Eq:diffusion_non_boost_invariant}.

\begin{figure}[htb]
  \vspace{0.4cm}
  \includegraphics[width=0.48\textwidth]{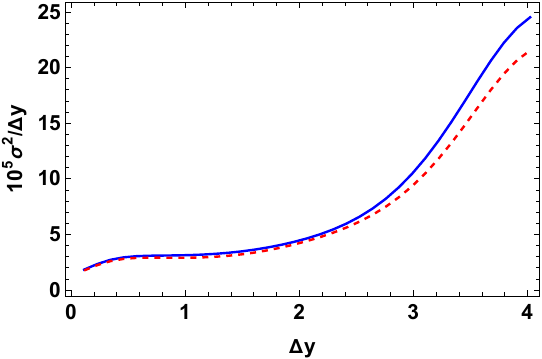}
  \caption{Rapidity dependence of the second order baryon density fluctuations at the freeze-out surface. Solid curves are the ones of boost-invariant limit Eq.\eqref{eq:linearEq}, dashed ones represent the ones within the non-boost-invariant fluid  (Eq.\eqref{Eq:diffusion_non_boost_invariant}).}
  \label{fig:C2FreezeOut}
\end{figure}

It requires the profile of the non-boost-invariant background to estimate such advection effects. As the profile of temperature and chemical potential in Fig.\ref{fig:Tmu}, the profile for $u^\tau$ and $\tau u^\eta$ are extracted from the hydrodynamic simulation of Au+Au collisions at 19.6GeV (\texttt{Scenario I}) and shown in Figs.\ref{fig:utau_ueta}. One can see the moderate acceleration of the non-boost-invariant effects ($\tau u^\eta$) at 19.6GeV. To see its consequence on the dynamics of the baryon density near the critical point, the advection-diffusion equation \eqref{Eq:diffusion_non_boost_invariant} is solved numerically as the way in Sec.\ref{sec:model} and shown in Fig.\ref{fig:C2FreezeOut}. For comparison, the diffusion equation in the boost-invariant limit \eqref{eq:linearEq} has also been studied. As shown in Fig.\ref{fig:C2FreezeOut}, the effects of the advection due to the non-boost-invariant flow acceleration are visible but moderate at 19.6GeV of the Au+Au collisions.

\bibliographystyle{apsrev4-2}
\bibliography{ModelBHydro}

\begin{thebibliography}{58}%
\makeatletter
\providecommand \@ifxundefined [1]{%
 \@ifx{#1\undefined}
}%
\providecommand \@ifnum [1]{%
 \ifnum #1\expandafter \@firstoftwo
 \else \expandafter \@secondoftwo
 \fi
}%
\providecommand \@ifx [1]{%
 \ifx #1\expandafter \@firstoftwo
 \else \expandafter \@secondoftwo
 \fi
}%
\providecommand \natexlab [1]{#1}%
\providecommand \enquote  [1]{``#1''}%
\providecommand \bibnamefont  [1]{#1}%
\providecommand \bibfnamefont [1]{#1}%
\providecommand \citenamefont [1]{#1}%
\providecommand \href@noop [0]{\@secondoftwo}%
\providecommand \href [0]{\begingroup \@sanitize@url \@href}%
\providecommand \@href[1]{\@@startlink{#1}\@@href}%
\providecommand \@@href[1]{\endgroup#1\@@endlink}%
\providecommand \@sanitize@url [0]{\catcode `\\12\catcode `\$12\catcode `\&12\catcode `\#12\catcode `\^12\catcode `\_12\catcode `\%12\relax}%
\providecommand \@@startlink[1]{}%
\providecommand \@@endlink[0]{}%
\providecommand \url  [0]{\begingroup\@sanitize@url \@url }%
\providecommand \@url [1]{\endgroup\@href {#1}{\urlprefix }}%
\providecommand \urlprefix  [0]{URL }%
\providecommand \Eprint [0]{\href }%
\providecommand \doibase [0]{https://doi.org/}%
\providecommand \selectlanguage [0]{\@gobble}%
\providecommand \bibinfo  [0]{\@secondoftwo}%
\providecommand \bibfield  [0]{\@secondoftwo}%
\providecommand \translation [1]{[#1]}%
\providecommand \BibitemOpen [0]{}%
\providecommand \bibitemStop [0]{}%
\providecommand \bibitemNoStop [0]{.\EOS\space}%
\providecommand \EOS [0]{\spacefactor3000\relax}%
\providecommand \BibitemShut  [1]{\csname bibitem#1\endcsname}%
\let\auto@bib@innerbib\@empty
\bibitem [{\citenamefont {Aoki}\ \emph {et~al.}(2006)\citenamefont {Aoki}, \citenamefont {Endrodi}, \citenamefont {Fodor}, \citenamefont {Katz},\ and\ \citenamefont {Szabo}}]{Aoki:2006we}%
  \BibitemOpen
  \bibfield  {author} {\bibinfo {author} {\bibfnamefont {Y.}~\bibnamefont {Aoki}}, \bibinfo {author} {\bibfnamefont {G.}~\bibnamefont {Endrodi}}, \bibinfo {author} {\bibfnamefont {Z.}~\bibnamefont {Fodor}}, \bibinfo {author} {\bibfnamefont {S.~D.}\ \bibnamefont {Katz}},\ and\ \bibinfo {author} {\bibfnamefont {K.~K.}\ \bibnamefont {Szabo}},\ }\href {https://doi.org/10.1038/nature05120} {\bibfield  {journal} {\bibinfo  {journal} {Nature}\ }\textbf {\bibinfo {volume} {443}},\ \bibinfo {pages} {675} (\bibinfo {year} {2006})},\ \Eprint {https://arxiv.org/abs/hep-lat/0611014} {arXiv:hep-lat/0611014} \BibitemShut {NoStop}%
\bibitem [{\citenamefont {Ding}\ \emph {et~al.}(2015)\citenamefont {Ding}, \citenamefont {Karsch},\ and\ \citenamefont {Mukherjee}}]{Ding:2015ona}%
  \BibitemOpen
  \bibfield  {author} {\bibinfo {author} {\bibfnamefont {H.-T.}\ \bibnamefont {Ding}}, \bibinfo {author} {\bibfnamefont {F.}~\bibnamefont {Karsch}},\ and\ \bibinfo {author} {\bibfnamefont {S.}~\bibnamefont {Mukherjee}},\ }\href {https://doi.org/10.1142/S0218301315300076} {\bibfield  {journal} {\bibinfo  {journal} {Int. J. Mod. Phys. E}\ }\textbf {\bibinfo {volume} {24}},\ \bibinfo {pages} {1530007} (\bibinfo {year} {2015})},\ \Eprint {https://arxiv.org/abs/1504.05274} {arXiv:1504.05274 [hep-lat]} \BibitemShut {NoStop}%
\bibitem [{\citenamefont {Bazavov}\ \emph {et~al.}(2019)\citenamefont {Bazavov}, \citenamefont {Karsch}, \citenamefont {Mukherjee},\ and\ \citenamefont {Petreczky}}]{Bazavov:2019lgz}%
  \BibitemOpen
  \bibfield  {author} {\bibinfo {author} {\bibfnamefont {A.}~\bibnamefont {Bazavov}}, \bibinfo {author} {\bibfnamefont {F.}~\bibnamefont {Karsch}}, \bibinfo {author} {\bibfnamefont {S.}~\bibnamefont {Mukherjee}},\ and\ \bibinfo {author} {\bibfnamefont {P.}~\bibnamefont {Petreczky}} (\bibinfo {collaboration} {USQCD}),\ }\href {https://doi.org/10.1140/epja/i2019-12922-0} {\bibfield  {journal} {\bibinfo  {journal} {Eur. Phys. J. A}\ }\textbf {\bibinfo {volume} {55}},\ \bibinfo {pages} {194} (\bibinfo {year} {2019})},\ \Eprint {https://arxiv.org/abs/1904.09951} {arXiv:1904.09951 [hep-lat]} \BibitemShut {NoStop}%
\bibitem [{\citenamefont {Ratti}(2018)}]{Ratti:2018ksb}%
  \BibitemOpen
  \bibfield  {author} {\bibinfo {author} {\bibfnamefont {C.}~\bibnamefont {Ratti}},\ }\href {https://doi.org/10.1088/1361-6633/aabb97} {\bibfield  {journal} {\bibinfo  {journal} {Rept. Prog. Phys.}\ }\textbf {\bibinfo {volume} {81}},\ \bibinfo {pages} {084301} (\bibinfo {year} {2018})},\ \Eprint {https://arxiv.org/abs/1804.07810} {arXiv:1804.07810 [hep-lat]} \BibitemShut {NoStop}%
\bibitem [{\citenamefont {Fischer}(2019)}]{Fischer:2018sdj}%
  \BibitemOpen
  \bibfield  {author} {\bibinfo {author} {\bibfnamefont {C.~S.}\ \bibnamefont {Fischer}},\ }\href {https://doi.org/10.1016/j.ppnp.2019.01.002} {\bibfield  {journal} {\bibinfo  {journal} {Prog. Part. Nucl. Phys.}\ }\textbf {\bibinfo {volume} {105}},\ \bibinfo {pages} {1} (\bibinfo {year} {2019})},\ \Eprint {https://arxiv.org/abs/1810.12938} {arXiv:1810.12938 [hep-ph]} \BibitemShut {NoStop}%
\bibitem [{\citenamefont {Fukushima}\ and\ \citenamefont {Hatsuda}(2011)}]{Fukushima:2010bq}%
  \BibitemOpen
  \bibfield  {author} {\bibinfo {author} {\bibfnamefont {K.}~\bibnamefont {Fukushima}}\ and\ \bibinfo {author} {\bibfnamefont {T.}~\bibnamefont {Hatsuda}},\ }\href {https://doi.org/10.1088/0034-4885/74/1/014001} {\bibfield  {journal} {\bibinfo  {journal} {Rept. Prog. Phys.}\ }\textbf {\bibinfo {volume} {74}},\ \bibinfo {pages} {014001} (\bibinfo {year} {2011})},\ \Eprint {https://arxiv.org/abs/1005.4814} {arXiv:1005.4814 [hep-ph]} \BibitemShut {NoStop}%
\bibitem [{\citenamefont {Fukushima}\ and\ \citenamefont {Sasaki}(2013)}]{Fukushima:2013rx}%
  \BibitemOpen
  \bibfield  {author} {\bibinfo {author} {\bibfnamefont {K.}~\bibnamefont {Fukushima}}\ and\ \bibinfo {author} {\bibfnamefont {C.}~\bibnamefont {Sasaki}},\ }\href {https://doi.org/10.1016/j.ppnp.2013.05.003} {\bibfield  {journal} {\bibinfo  {journal} {Prog. Part. Nucl. Phys.}\ }\textbf {\bibinfo {volume} {72}},\ \bibinfo {pages} {99} (\bibinfo {year} {2013})},\ \Eprint {https://arxiv.org/abs/1301.6377} {arXiv:1301.6377 [hep-ph]} \BibitemShut {NoStop}%
\bibitem [{\citenamefont {Fu}(2022)}]{Fu:2022gou}%
  \BibitemOpen
  \bibfield  {author} {\bibinfo {author} {\bibfnamefont {W.-j.}\ \bibnamefont {Fu}},\ }\href {https://doi.org/10.1088/1572-9494/ac86be} {\bibfield  {journal} {\bibinfo  {journal} {Commun. Theor. Phys.}\ }\textbf {\bibinfo {volume} {74}},\ \bibinfo {pages} {097304} (\bibinfo {year} {2022})},\ \Eprint {https://arxiv.org/abs/2205.00468} {arXiv:2205.00468 [hep-ph]} \BibitemShut {NoStop}%
\bibitem [{\citenamefont {Stephanov}(2011)}]{Stephanov:2011pb}%
  \BibitemOpen
  \bibfield  {author} {\bibinfo {author} {\bibfnamefont {M.~A.}\ \bibnamefont {Stephanov}},\ }\href {https://doi.org/10.1103/PhysRevLett.107.052301} {\bibfield  {journal} {\bibinfo  {journal} {Phys. Rev. Lett.}\ }\textbf {\bibinfo {volume} {107}},\ \bibinfo {pages} {052301} (\bibinfo {year} {2011})},\ \Eprint {https://arxiv.org/abs/1104.1627} {arXiv:1104.1627 [hep-ph]} \BibitemShut {NoStop}%
\bibitem [{\citenamefont {Athanasiou}\ \emph {et~al.}(2010)\citenamefont {Athanasiou}, \citenamefont {Rajagopal},\ and\ \citenamefont {Stephanov}}]{Athanasiou:2010kw}%
  \BibitemOpen
  \bibfield  {author} {\bibinfo {author} {\bibfnamefont {C.}~\bibnamefont {Athanasiou}}, \bibinfo {author} {\bibfnamefont {K.}~\bibnamefont {Rajagopal}},\ and\ \bibinfo {author} {\bibfnamefont {M.}~\bibnamefont {Stephanov}},\ }\href {https://doi.org/10.1103/PhysRevD.82.074008} {\bibfield  {journal} {\bibinfo  {journal} {Phys. Rev. D}\ }\textbf {\bibinfo {volume} {82}},\ \bibinfo {pages} {074008} (\bibinfo {year} {2010})},\ \Eprint {https://arxiv.org/abs/1006.4636} {arXiv:1006.4636 [hep-ph]} \BibitemShut {NoStop}%
\bibitem [{\citenamefont {Adam}\ \emph {et~al.}(2021)\citenamefont {Adam} \emph {et~al.}}]{STAR:2020tga}%
  \BibitemOpen
  \bibfield  {author} {\bibinfo {author} {\bibfnamefont {J.}~\bibnamefont {Adam}} \emph {et~al.} (\bibinfo {collaboration} {STAR}),\ }\href {https://doi.org/10.1103/PhysRevLett.126.092301} {\bibfield  {journal} {\bibinfo  {journal} {Phys. Rev. Lett.}\ }\textbf {\bibinfo {volume} {126}},\ \bibinfo {pages} {092301} (\bibinfo {year} {2021})},\ \Eprint {https://arxiv.org/abs/2001.02852} {arXiv:2001.02852 [nucl-ex]} \BibitemShut {NoStop}%
\bibitem [{\citenamefont {Abdallah}\ \emph {et~al.}(2021)\citenamefont {Abdallah} \emph {et~al.}}]{STAR:2021iop}%
  \BibitemOpen
  \bibfield  {author} {\bibinfo {author} {\bibfnamefont {M.}~\bibnamefont {Abdallah}} \emph {et~al.} (\bibinfo {collaboration} {STAR}),\ }\href {https://doi.org/10.1103/PhysRevC.104.024902} {\bibfield  {journal} {\bibinfo  {journal} {Phys. Rev. C}\ }\textbf {\bibinfo {volume} {104}},\ \bibinfo {pages} {024902} (\bibinfo {year} {2021})},\ \Eprint {https://arxiv.org/abs/2101.12413} {arXiv:2101.12413 [nucl-ex]} \BibitemShut {NoStop}%
\bibitem [{\citenamefont {Asakawa}\ and\ \citenamefont {Kitazawa}(2016)}]{Asakawa:2015ybt}%
  \BibitemOpen
  \bibfield  {author} {\bibinfo {author} {\bibfnamefont {M.}~\bibnamefont {Asakawa}}\ and\ \bibinfo {author} {\bibfnamefont {M.}~\bibnamefont {Kitazawa}},\ }\href {https://doi.org/10.1016/j.ppnp.2016.04.002} {\bibfield  {journal} {\bibinfo  {journal} {Prog. Part. Nucl. Phys.}\ }\textbf {\bibinfo {volume} {90}},\ \bibinfo {pages} {299} (\bibinfo {year} {2016})},\ \Eprint {https://arxiv.org/abs/1512.05038} {arXiv:1512.05038 [nucl-th]} \BibitemShut {NoStop}%
\bibitem [{\citenamefont {Bzdak}\ \emph {et~al.}(2020)\citenamefont {Bzdak}, \citenamefont {Esumi}, \citenamefont {Koch}, \citenamefont {Liao}, \citenamefont {Stephanov},\ and\ \citenamefont {Xu}}]{Bzdak:2019pkr}%
  \BibitemOpen
  \bibfield  {author} {\bibinfo {author} {\bibfnamefont {A.}~\bibnamefont {Bzdak}}, \bibinfo {author} {\bibfnamefont {S.}~\bibnamefont {Esumi}}, \bibinfo {author} {\bibfnamefont {V.}~\bibnamefont {Koch}}, \bibinfo {author} {\bibfnamefont {J.}~\bibnamefont {Liao}}, \bibinfo {author} {\bibfnamefont {M.}~\bibnamefont {Stephanov}},\ and\ \bibinfo {author} {\bibfnamefont {N.}~\bibnamefont {Xu}},\ }\href {https://doi.org/10.1016/j.physrep.2020.01.005} {\bibfield  {journal} {\bibinfo  {journal} {Phys. Rept.}\ }\textbf {\bibinfo {volume} {853}},\ \bibinfo {pages} {1} (\bibinfo {year} {2020})},\ \Eprint {https://arxiv.org/abs/1906.00936} {arXiv:1906.00936 [nucl-th]} \BibitemShut {NoStop}%
\bibitem [{\citenamefont {Bluhm}\ \emph {et~al.}(2020)\citenamefont {Bluhm} \emph {et~al.}}]{Bluhm:2020mpc}%
  \BibitemOpen
  \bibfield  {author} {\bibinfo {author} {\bibfnamefont {M.}~\bibnamefont {Bluhm}} \emph {et~al.},\ }\href {https://doi.org/10.1016/j.nuclphysa.2020.122016} {\bibfield  {journal} {\bibinfo  {journal} {Nucl. Phys. A}\ }\textbf {\bibinfo {volume} {1003}},\ \bibinfo {pages} {122016} (\bibinfo {year} {2020})},\ \Eprint {https://arxiv.org/abs/2001.08831} {arXiv:2001.08831 [nucl-th]} \BibitemShut {NoStop}%
\bibitem [{\citenamefont {Wu}\ \emph {et~al.}(2021)\citenamefont {Wu}, \citenamefont {Shen},\ and\ \citenamefont {Song}}]{Wu:2021xgu}%
  \BibitemOpen
  \bibfield  {author} {\bibinfo {author} {\bibfnamefont {S.}~\bibnamefont {Wu}}, \bibinfo {author} {\bibfnamefont {C.}~\bibnamefont {Shen}},\ and\ \bibinfo {author} {\bibfnamefont {H.}~\bibnamefont {Song}},\ }\href {https://doi.org/10.1088/0256-307X/38/8/081201} {\bibfield  {journal} {\bibinfo  {journal} {Chin. Phys. Lett.}\ }\textbf {\bibinfo {volume} {38}},\ \bibinfo {pages} {081201} (\bibinfo {year} {2021})},\ \Eprint {https://arxiv.org/abs/2104.13250} {arXiv:2104.13250 [nucl-th]} \BibitemShut {NoStop}%
\bibitem [{\citenamefont {An}\ \emph {et~al.}(2022)\citenamefont {An} \emph {et~al.}}]{An:2021wof}%
  \BibitemOpen
  \bibfield  {author} {\bibinfo {author} {\bibfnamefont {X.}~\bibnamefont {An}} \emph {et~al.},\ }\href {https://doi.org/10.1016/j.nuclphysa.2021.122343} {\bibfield  {journal} {\bibinfo  {journal} {Nucl. Phys. A}\ }\textbf {\bibinfo {volume} {1017}},\ \bibinfo {pages} {122343} (\bibinfo {year} {2022})},\ \Eprint {https://arxiv.org/abs/2108.13867} {arXiv:2108.13867 [nucl-th]} \BibitemShut {NoStop}%
\bibitem [{\citenamefont {Du}\ \emph {et~al.}(2024{\natexlab{a}})\citenamefont {Du}, \citenamefont {Sorensen},\ and\ \citenamefont {Stephanov}}]{Du:2024wjm}%
  \BibitemOpen
  \bibfield  {author} {\bibinfo {author} {\bibfnamefont {L.}~\bibnamefont {Du}}, \bibinfo {author} {\bibfnamefont {A.}~\bibnamefont {Sorensen}},\ and\ \bibinfo {author} {\bibfnamefont {M.}~\bibnamefont {Stephanov}}\ }(\bibinfo {year} {2024})\ \Eprint {https://arxiv.org/abs/2402.10183} {arXiv:2402.10183 [nucl-th]} \BibitemShut {NoStop}%
\bibitem [{\citenamefont {Berdnikov}\ and\ \citenamefont {Rajagopal}(2000)}]{Berdnikov:1999ph}%
  \BibitemOpen
  \bibfield  {author} {\bibinfo {author} {\bibfnamefont {B.}~\bibnamefont {Berdnikov}}\ and\ \bibinfo {author} {\bibfnamefont {K.}~\bibnamefont {Rajagopal}},\ }\href {https://doi.org/10.1103/PhysRevD.61.105017} {\bibfield  {journal} {\bibinfo  {journal} {Phys. Rev. D}\ }\textbf {\bibinfo {volume} {61}},\ \bibinfo {pages} {105017} (\bibinfo {year} {2000})},\ \Eprint {https://arxiv.org/abs/hep-ph/9912274} {arXiv:hep-ph/9912274} \BibitemShut {NoStop}%
\bibitem [{\citenamefont {Nonaka}\ and\ \citenamefont {Asakawa}(2005)}]{Nonaka:2004pg}%
  \BibitemOpen
  \bibfield  {author} {\bibinfo {author} {\bibfnamefont {C.}~\bibnamefont {Nonaka}}\ and\ \bibinfo {author} {\bibfnamefont {M.}~\bibnamefont {Asakawa}},\ }\href {https://doi.org/10.1103/PhysRevC.71.044904} {\bibfield  {journal} {\bibinfo  {journal} {Phys. Rev. C}\ }\textbf {\bibinfo {volume} {71}},\ \bibinfo {pages} {044904} (\bibinfo {year} {2005})},\ \Eprint {https://arxiv.org/abs/nucl-th/0410078} {arXiv:nucl-th/0410078} \BibitemShut {NoStop}%
\bibitem [{\citenamefont {Mukherjee}\ \emph {et~al.}(2015)\citenamefont {Mukherjee}, \citenamefont {Venugopalan},\ and\ \citenamefont {Yin}}]{Mukherjee:2015swa}%
  \BibitemOpen
  \bibfield  {author} {\bibinfo {author} {\bibfnamefont {S.}~\bibnamefont {Mukherjee}}, \bibinfo {author} {\bibfnamefont {R.}~\bibnamefont {Venugopalan}},\ and\ \bibinfo {author} {\bibfnamefont {Y.}~\bibnamefont {Yin}},\ }\href {https://doi.org/10.1103/PhysRevC.92.034912} {\bibfield  {journal} {\bibinfo  {journal} {Phys. Rev. C}\ }\textbf {\bibinfo {volume} {92}},\ \bibinfo {pages} {034912} (\bibinfo {year} {2015})},\ \Eprint {https://arxiv.org/abs/1506.00645} {arXiv:1506.00645 [hep-ph]} \BibitemShut {NoStop}%
\bibitem [{\citenamefont {Tang}\ \emph {et~al.}(2023)\citenamefont {Tang}, \citenamefont {Wu},\ and\ \citenamefont {Song}}]{Tang:2023zvj}%
  \BibitemOpen
  \bibfield  {author} {\bibinfo {author} {\bibfnamefont {S.}~\bibnamefont {Tang}}, \bibinfo {author} {\bibfnamefont {S.}~\bibnamefont {Wu}},\ and\ \bibinfo {author} {\bibfnamefont {H.}~\bibnamefont {Song}},\ }\href {https://doi.org/10.1103/PhysRevC.108.034901} {\bibfield  {journal} {\bibinfo  {journal} {Phys. Rev. C}\ }\textbf {\bibinfo {volume} {108}},\ \bibinfo {pages} {034901} (\bibinfo {year} {2023})},\ \Eprint {https://arxiv.org/abs/2303.15017} {arXiv:2303.15017 [nucl-th]} \BibitemShut {NoStop}%
\bibitem [{\citenamefont {Nahrgang}\ \emph {et~al.}(2019)\citenamefont {Nahrgang}, \citenamefont {Bluhm}, \citenamefont {Schaefer},\ and\ \citenamefont {Bass}}]{Nahrgang:2018afz}%
  \BibitemOpen
  \bibfield  {author} {\bibinfo {author} {\bibfnamefont {M.}~\bibnamefont {Nahrgang}}, \bibinfo {author} {\bibfnamefont {M.}~\bibnamefont {Bluhm}}, \bibinfo {author} {\bibfnamefont {T.}~\bibnamefont {Schaefer}},\ and\ \bibinfo {author} {\bibfnamefont {S.~A.}\ \bibnamefont {Bass}},\ }\href {https://doi.org/10.1103/PhysRevD.99.116015} {\bibfield  {journal} {\bibinfo  {journal} {Phys. Rev. D}\ }\textbf {\bibinfo {volume} {99}},\ \bibinfo {pages} {116015} (\bibinfo {year} {2019})},\ \Eprint {https://arxiv.org/abs/1804.05728} {arXiv:1804.05728 [nucl-th]} \BibitemShut {NoStop}%
\bibitem [{\citenamefont {Nahrgang}\ and\ \citenamefont {Bluhm}(2020)}]{Nahrgang:2020yxm}%
  \BibitemOpen
  \bibfield  {author} {\bibinfo {author} {\bibfnamefont {M.}~\bibnamefont {Nahrgang}}\ and\ \bibinfo {author} {\bibfnamefont {M.}~\bibnamefont {Bluhm}},\ }\href {https://doi.org/10.1103/PhysRevD.102.094017} {\bibfield  {journal} {\bibinfo  {journal} {Phys. Rev. D}\ }\textbf {\bibinfo {volume} {102}},\ \bibinfo {pages} {094017} (\bibinfo {year} {2020})},\ \Eprint {https://arxiv.org/abs/2007.10371} {arXiv:2007.10371 [nucl-th]} \BibitemShut {NoStop}%
\bibitem [{\citenamefont {Sakaida}\ \emph {et~al.}(2017)\citenamefont {Sakaida}, \citenamefont {Asakawa}, \citenamefont {Fujii},\ and\ \citenamefont {Kitazawa}}]{Sakaida:2017rtj}%
  \BibitemOpen
  \bibfield  {author} {\bibinfo {author} {\bibfnamefont {M.}~\bibnamefont {Sakaida}}, \bibinfo {author} {\bibfnamefont {M.}~\bibnamefont {Asakawa}}, \bibinfo {author} {\bibfnamefont {H.}~\bibnamefont {Fujii}},\ and\ \bibinfo {author} {\bibfnamefont {M.}~\bibnamefont {Kitazawa}},\ }\href {https://doi.org/10.1103/PhysRevC.95.064905} {\bibfield  {journal} {\bibinfo  {journal} {Phys. Rev. C}\ }\textbf {\bibinfo {volume} {95}},\ \bibinfo {pages} {064905} (\bibinfo {year} {2017})},\ \Eprint {https://arxiv.org/abs/1703.08008} {arXiv:1703.08008 [nucl-th]} \BibitemShut {NoStop}%
\bibitem [{\citenamefont {Pihan}\ \emph {et~al.}(2023)\citenamefont {Pihan}, \citenamefont {Bluhm}, \citenamefont {Kitazawa}, \citenamefont {Sami},\ and\ \citenamefont {Nahrgang}}]{Pihan:2022xcl}%
  \BibitemOpen
  \bibfield  {author} {\bibinfo {author} {\bibfnamefont {G.}~\bibnamefont {Pihan}}, \bibinfo {author} {\bibfnamefont {M.}~\bibnamefont {Bluhm}}, \bibinfo {author} {\bibfnamefont {M.}~\bibnamefont {Kitazawa}}, \bibinfo {author} {\bibfnamefont {T.}~\bibnamefont {Sami}},\ and\ \bibinfo {author} {\bibfnamefont {M.}~\bibnamefont {Nahrgang}},\ }\href {https://doi.org/10.1103/PhysRevC.107.014908} {\bibfield  {journal} {\bibinfo  {journal} {Phys. Rev. C}\ }\textbf {\bibinfo {volume} {107}},\ \bibinfo {pages} {014908} (\bibinfo {year} {2023})},\ \Eprint {https://arxiv.org/abs/2205.12834} {arXiv:2205.12834 [nucl-th]} \BibitemShut {NoStop}%
\bibitem [{\citenamefont {Nahrgang}\ \emph {et~al.}(2011)\citenamefont {Nahrgang}, \citenamefont {Leupold}, \citenamefont {Herold},\ and\ \citenamefont {Bleicher}}]{Nahrgang:2011mg}%
  \BibitemOpen
  \bibfield  {author} {\bibinfo {author} {\bibfnamefont {M.}~\bibnamefont {Nahrgang}}, \bibinfo {author} {\bibfnamefont {S.}~\bibnamefont {Leupold}}, \bibinfo {author} {\bibfnamefont {C.}~\bibnamefont {Herold}},\ and\ \bibinfo {author} {\bibfnamefont {M.}~\bibnamefont {Bleicher}},\ }\href {https://doi.org/10.1103/PhysRevC.84.024912} {\bibfield  {journal} {\bibinfo  {journal} {Phys. Rev. C}\ }\textbf {\bibinfo {volume} {84}},\ \bibinfo {pages} {024912} (\bibinfo {year} {2011})},\ \Eprint {https://arxiv.org/abs/1105.0622} {arXiv:1105.0622 [nucl-th]} \BibitemShut {NoStop}%
\bibitem [{\citenamefont {Nahrgang}\ \emph {et~al.}(2013)\citenamefont {Nahrgang}, \citenamefont {Herold}, \citenamefont {Leupold}, \citenamefont {Mishustin},\ and\ \citenamefont {Bleicher}}]{Nahrgang:2011vn}%
  \BibitemOpen
  \bibfield  {author} {\bibinfo {author} {\bibfnamefont {M.}~\bibnamefont {Nahrgang}}, \bibinfo {author} {\bibfnamefont {C.}~\bibnamefont {Herold}}, \bibinfo {author} {\bibfnamefont {S.}~\bibnamefont {Leupold}}, \bibinfo {author} {\bibfnamefont {I.}~\bibnamefont {Mishustin}},\ and\ \bibinfo {author} {\bibfnamefont {M.}~\bibnamefont {Bleicher}},\ }\href {https://doi.org/10.1088/0954-3899/40/5/055108} {\bibfield  {journal} {\bibinfo  {journal} {J. Phys. G}\ }\textbf {\bibinfo {volume} {40}},\ \bibinfo {pages} {055108} (\bibinfo {year} {2013})},\ \Eprint {https://arxiv.org/abs/1105.1962} {arXiv:1105.1962 [nucl-th]} \BibitemShut {NoStop}%
\bibitem [{\citenamefont {Herold}\ \emph {et~al.}(2014)\citenamefont {Herold}, \citenamefont {Nahrgang}, \citenamefont {Yan},\ and\ \citenamefont {Kobdaj}}]{Herold:2014zoa}%
  \BibitemOpen
  \bibfield  {author} {\bibinfo {author} {\bibfnamefont {C.}~\bibnamefont {Herold}}, \bibinfo {author} {\bibfnamefont {M.}~\bibnamefont {Nahrgang}}, \bibinfo {author} {\bibfnamefont {Y.}~\bibnamefont {Yan}},\ and\ \bibinfo {author} {\bibfnamefont {C.}~\bibnamefont {Kobdaj}},\ }\href {https://doi.org/10.1088/0954-3899/41/11/115106} {\bibfield  {journal} {\bibinfo  {journal} {J. Phys. G}\ }\textbf {\bibinfo {volume} {41}},\ \bibinfo {pages} {115106} (\bibinfo {year} {2014})},\ \Eprint {https://arxiv.org/abs/1407.8277} {arXiv:1407.8277 [hep-ph]} \BibitemShut {NoStop}%
\bibitem [{\citenamefont {Kapusta}\ \emph {et~al.}(2012)\citenamefont {Kapusta}, \citenamefont {Muller},\ and\ \citenamefont {Stephanov}}]{Kapusta:2011gt}%
  \BibitemOpen
  \bibfield  {author} {\bibinfo {author} {\bibfnamefont {J.~I.}\ \bibnamefont {Kapusta}}, \bibinfo {author} {\bibfnamefont {B.}~\bibnamefont {Muller}},\ and\ \bibinfo {author} {\bibfnamefont {M.}~\bibnamefont {Stephanov}},\ }\href {https://doi.org/10.1103/PhysRevC.85.054906} {\bibfield  {journal} {\bibinfo  {journal} {Phys. Rev. C}\ }\textbf {\bibinfo {volume} {85}},\ \bibinfo {pages} {054906} (\bibinfo {year} {2012})},\ \Eprint {https://arxiv.org/abs/1112.6405} {arXiv:1112.6405 [nucl-th]} \BibitemShut {NoStop}%
\bibitem [{\citenamefont {An}\ \emph {et~al.}(2020)\citenamefont {An}, \citenamefont {Ba\c{s}ar}, \citenamefont {Stephanov},\ and\ \citenamefont {Yee}}]{An:2019csj}%
  \BibitemOpen
  \bibfield  {author} {\bibinfo {author} {\bibfnamefont {X.}~\bibnamefont {An}}, \bibinfo {author} {\bibfnamefont {G.}~\bibnamefont {Ba\c{s}ar}}, \bibinfo {author} {\bibfnamefont {M.}~\bibnamefont {Stephanov}},\ and\ \bibinfo {author} {\bibfnamefont {H.-U.}\ \bibnamefont {Yee}},\ }\href {https://doi.org/10.1103/PhysRevC.102.034901} {\bibfield  {journal} {\bibinfo  {journal} {Phys. Rev. C}\ }\textbf {\bibinfo {volume} {102}},\ \bibinfo {pages} {034901} (\bibinfo {year} {2020})},\ \Eprint {https://arxiv.org/abs/1912.13456} {arXiv:1912.13456 [hep-th]} \BibitemShut {NoStop}%
\bibitem [{\citenamefont {An}\ \emph {et~al.}(2021)\citenamefont {An}, \citenamefont {Ba\c{s}ar}, \citenamefont {Stephanov},\ and\ \citenamefont {Yee}}]{An:2020vri}%
  \BibitemOpen
  \bibfield  {author} {\bibinfo {author} {\bibfnamefont {X.}~\bibnamefont {An}}, \bibinfo {author} {\bibfnamefont {G.}~\bibnamefont {Ba\c{s}ar}}, \bibinfo {author} {\bibfnamefont {M.}~\bibnamefont {Stephanov}},\ and\ \bibinfo {author} {\bibfnamefont {H.-U.}\ \bibnamefont {Yee}},\ }\href {https://doi.org/10.1103/PhysRevLett.127.072301} {\bibfield  {journal} {\bibinfo  {journal} {Phys. Rev. Lett.}\ }\textbf {\bibinfo {volume} {127}},\ \bibinfo {pages} {072301} (\bibinfo {year} {2021})},\ \Eprint {https://arxiv.org/abs/2009.10742} {arXiv:2009.10742 [hep-th]} \BibitemShut {NoStop}%
\bibitem [{\citenamefont {Stephanov}\ and\ \citenamefont {Yin}(2018)}]{Stephanov:2017ghc}%
  \BibitemOpen
  \bibfield  {author} {\bibinfo {author} {\bibfnamefont {M.}~\bibnamefont {Stephanov}}\ and\ \bibinfo {author} {\bibfnamefont {Y.}~\bibnamefont {Yin}},\ }\href {https://doi.org/10.1103/PhysRevD.98.036006} {\bibfield  {journal} {\bibinfo  {journal} {Phys. Rev. D}\ }\textbf {\bibinfo {volume} {98}},\ \bibinfo {pages} {036006} (\bibinfo {year} {2018})},\ \Eprint {https://arxiv.org/abs/1712.10305} {arXiv:1712.10305 [nucl-th]} \BibitemShut {NoStop}%
\bibitem [{\citenamefont {Shen}\ and\ \citenamefont {Schenke}(2018)}]{Shen:2017bsr}%
  \BibitemOpen
  \bibfield  {author} {\bibinfo {author} {\bibfnamefont {C.}~\bibnamefont {Shen}}\ and\ \bibinfo {author} {\bibfnamefont {B.}~\bibnamefont {Schenke}},\ }\href {https://doi.org/10.1103/PhysRevC.97.024907} {\bibfield  {journal} {\bibinfo  {journal} {Phys. Rev. C}\ }\textbf {\bibinfo {volume} {97}},\ \bibinfo {pages} {024907} (\bibinfo {year} {2018})},\ \Eprint {https://arxiv.org/abs/1710.00881} {arXiv:1710.00881 [nucl-th]} \BibitemShut {NoStop}%
\bibitem [{\citenamefont {Du}\ \emph {et~al.}(2024{\natexlab{b}})\citenamefont {Du}, \citenamefont {Gao}, \citenamefont {Jeon},\ and\ \citenamefont {Gale}}]{Du:2023gnv}%
  \BibitemOpen
  \bibfield  {author} {\bibinfo {author} {\bibfnamefont {L.}~\bibnamefont {Du}}, \bibinfo {author} {\bibfnamefont {H.}~\bibnamefont {Gao}}, \bibinfo {author} {\bibfnamefont {S.}~\bibnamefont {Jeon}},\ and\ \bibinfo {author} {\bibfnamefont {C.}~\bibnamefont {Gale}},\ }\href {https://doi.org/10.1103/PhysRevC.109.014907} {\bibfield  {journal} {\bibinfo  {journal} {Phys. Rev. C}\ }\textbf {\bibinfo {volume} {109}},\ \bibinfo {pages} {014907} (\bibinfo {year} {2024}{\natexlab{b}})},\ \Eprint {https://arxiv.org/abs/2302.13852} {arXiv:2302.13852 [nucl-th]} \BibitemShut {NoStop}%
\bibitem [{\citenamefont {Du}\ \emph {et~al.}(2021)\citenamefont {Du}, \citenamefont {An},\ and\ \citenamefont {Heinz}}]{Du:2021zqz}%
  \BibitemOpen
  \bibfield  {author} {\bibinfo {author} {\bibfnamefont {L.}~\bibnamefont {Du}}, \bibinfo {author} {\bibfnamefont {X.}~\bibnamefont {An}},\ and\ \bibinfo {author} {\bibfnamefont {U.}~\bibnamefont {Heinz}},\ }\href {https://doi.org/10.1103/PhysRevC.104.064904} {\bibfield  {journal} {\bibinfo  {journal} {Phys. Rev. C}\ }\textbf {\bibinfo {volume} {104}},\ \bibinfo {pages} {064904} (\bibinfo {year} {2021})},\ \Eprint {https://arxiv.org/abs/2107.02302} {arXiv:2107.02302 [hep-ph]} \BibitemShut {NoStop}%
\bibitem [{\citenamefont {Brewer}\ \emph {et~al.}(2018)\citenamefont {Brewer}, \citenamefont {Mukherjee}, \citenamefont {Rajagopal},\ and\ \citenamefont {Yin}}]{Brewer:2018abr}%
  \BibitemOpen
  \bibfield  {author} {\bibinfo {author} {\bibfnamefont {J.}~\bibnamefont {Brewer}}, \bibinfo {author} {\bibfnamefont {S.}~\bibnamefont {Mukherjee}}, \bibinfo {author} {\bibfnamefont {K.}~\bibnamefont {Rajagopal}},\ and\ \bibinfo {author} {\bibfnamefont {Y.}~\bibnamefont {Yin}},\ }\href {https://doi.org/10.1103/PhysRevC.98.061901} {\bibfield  {journal} {\bibinfo  {journal} {Phys. Rev. C}\ }\textbf {\bibinfo {volume} {98}},\ \bibinfo {pages} {061901} (\bibinfo {year} {2018})},\ \Eprint {https://arxiv.org/abs/1804.10215} {arXiv:1804.10215 [hep-ph]} \BibitemShut {NoStop}%
\bibitem [{\citenamefont {Son}\ and\ \citenamefont {Stephanov}(2004)}]{Son:2004iv}%
  \BibitemOpen
  \bibfield  {author} {\bibinfo {author} {\bibfnamefont {D.~T.}\ \bibnamefont {Son}}\ and\ \bibinfo {author} {\bibfnamefont {M.~A.}\ \bibnamefont {Stephanov}},\ }\href {https://doi.org/10.1103/PhysRevD.70.056001} {\bibfield  {journal} {\bibinfo  {journal} {Phys. Rev. D}\ }\textbf {\bibinfo {volume} {70}},\ \bibinfo {pages} {056001} (\bibinfo {year} {2004})},\ \Eprint {https://arxiv.org/abs/hep-ph/0401052} {arXiv:hep-ph/0401052} \BibitemShut {NoStop}%
\bibitem [{\citenamefont {Ling}\ \emph {et~al.}(2014)\citenamefont {Ling}, \citenamefont {Springer},\ and\ \citenamefont {Stephanov}}]{Ling:2013ksb}%
  \BibitemOpen
  \bibfield  {author} {\bibinfo {author} {\bibfnamefont {B.}~\bibnamefont {Ling}}, \bibinfo {author} {\bibfnamefont {T.}~\bibnamefont {Springer}},\ and\ \bibinfo {author} {\bibfnamefont {M.}~\bibnamefont {Stephanov}},\ }\href {https://doi.org/10.1103/PhysRevC.89.064901} {\bibfield  {journal} {\bibinfo  {journal} {Phys. Rev. C}\ }\textbf {\bibinfo {volume} {89}},\ \bibinfo {pages} {064901} (\bibinfo {year} {2014})},\ \Eprint {https://arxiv.org/abs/1310.6036} {arXiv:1310.6036 [nucl-th]} \BibitemShut {NoStop}%
\bibitem [{\citenamefont {Chao}\ and\ \citenamefont {Schaefer}(2021)}]{Chao:2020kcf}%
  \BibitemOpen
  \bibfield  {author} {\bibinfo {author} {\bibfnamefont {J.}~\bibnamefont {Chao}}\ and\ \bibinfo {author} {\bibfnamefont {T.}~\bibnamefont {Schaefer}},\ }\href {https://doi.org/10.1007/JHEP01(2021)071} {\bibfield  {journal} {\bibinfo  {journal} {JHEP}\ }\textbf {\bibinfo {volume} {01}},\ \bibinfo {pages} {071}},\ \Eprint {https://arxiv.org/abs/2008.01269} {arXiv:2008.01269 [hep-th]} \BibitemShut {NoStop}%
\bibitem [{\citenamefont {Chao}\ and\ \citenamefont {Schaefer}(2023)}]{Chao:2023kvz}%
  \BibitemOpen
  \bibfield  {author} {\bibinfo {author} {\bibfnamefont {J.}~\bibnamefont {Chao}}\ and\ \bibinfo {author} {\bibfnamefont {T.}~\bibnamefont {Schaefer}},\ }\href {https://doi.org/10.1007/JHEP06(2023)057} {\bibfield  {journal} {\bibinfo  {journal} {JHEP}\ }\textbf {\bibinfo {volume} {06}},\ \bibinfo {pages} {057}},\ \Eprint {https://arxiv.org/abs/2302.00720} {arXiv:2302.00720 [hep-ph]} \BibitemShut {NoStop}%
\bibitem [{\citenamefont {{Hu}}(2024)}]{2024arXiv240315825H}%
  \BibitemOpen
  \bibfield  {author} {\bibinfo {author} {\bibfnamefont {J.}~\bibnamefont {{Hu}}},\ }\href {https://doi.org/10.48550/arXiv.2403.15825} {\bibfield  {journal} {\bibinfo  {journal} {arXiv e-prints}\ ,\ \bibinfo {eid} {arXiv:2403.15825}} (\bibinfo {year} {2024})},\ \Eprint {https://arxiv.org/abs/2403.15825} {arXiv:2403.15825 [cond-mat.stat-mech]} \BibitemShut {NoStop}%
\bibitem [{\citenamefont {Motornenko}\ \emph {et~al.}(2020)\citenamefont {Motornenko}, \citenamefont {Steinheimer}, \citenamefont {Vovchenko}, \citenamefont {Schramm},\ and\ \citenamefont {Stoecker}}]{Motornenko:2019arp}%
  \BibitemOpen
  \bibfield  {author} {\bibinfo {author} {\bibfnamefont {A.}~\bibnamefont {Motornenko}}, \bibinfo {author} {\bibfnamefont {J.}~\bibnamefont {Steinheimer}}, \bibinfo {author} {\bibfnamefont {V.}~\bibnamefont {Vovchenko}}, \bibinfo {author} {\bibfnamefont {S.}~\bibnamefont {Schramm}},\ and\ \bibinfo {author} {\bibfnamefont {H.}~\bibnamefont {Stoecker}},\ }\href {https://doi.org/10.1103/PhysRevC.101.034904} {\bibfield  {journal} {\bibinfo  {journal} {Phys. Rev. C}\ }\textbf {\bibinfo {volume} {101}},\ \bibinfo {pages} {034904} (\bibinfo {year} {2020})},\ \Eprint {https://arxiv.org/abs/1905.00866} {arXiv:1905.00866 [hep-ph]} \BibitemShut {NoStop}%
\bibitem [{\citenamefont {Mukherjee}\ \emph {et~al.}(2017)\citenamefont {Mukherjee}, \citenamefont {Steinheimer},\ and\ \citenamefont {Schramm}}]{Mukherjee:2016nhb}%
  \BibitemOpen
  \bibfield  {author} {\bibinfo {author} {\bibfnamefont {A.}~\bibnamefont {Mukherjee}}, \bibinfo {author} {\bibfnamefont {J.}~\bibnamefont {Steinheimer}},\ and\ \bibinfo {author} {\bibfnamefont {S.}~\bibnamefont {Schramm}},\ }\href {https://doi.org/10.1103/PhysRevC.96.025205} {\bibfield  {journal} {\bibinfo  {journal} {Phys. Rev. C}\ }\textbf {\bibinfo {volume} {96}},\ \bibinfo {pages} {025205} (\bibinfo {year} {2017})},\ \Eprint {https://arxiv.org/abs/1611.10144} {arXiv:1611.10144 [nucl-th]} \BibitemShut {NoStop}%
\bibitem [{\citenamefont {Shen}\ \emph {et~al.}(2016)\citenamefont {Shen}, \citenamefont {Qiu}, \citenamefont {Song}, \citenamefont {Bernhard}, \citenamefont {Bass},\ and\ \citenamefont {Heinz}}]{Shen:2014vra}%
  \BibitemOpen
  \bibfield  {author} {\bibinfo {author} {\bibfnamefont {C.}~\bibnamefont {Shen}}, \bibinfo {author} {\bibfnamefont {Z.}~\bibnamefont {Qiu}}, \bibinfo {author} {\bibfnamefont {H.}~\bibnamefont {Song}}, \bibinfo {author} {\bibfnamefont {J.}~\bibnamefont {Bernhard}}, \bibinfo {author} {\bibfnamefont {S.}~\bibnamefont {Bass}},\ and\ \bibinfo {author} {\bibfnamefont {U.}~\bibnamefont {Heinz}},\ }\href {https://doi.org/10.1016/j.cpc.2015.08.039} {\bibfield  {journal} {\bibinfo  {journal} {Comput. Phys. Commun.}\ }\textbf {\bibinfo {volume} {199}},\ \bibinfo {pages} {61} (\bibinfo {year} {2016})},\ \Eprint {https://arxiv.org/abs/1409.8164} {arXiv:1409.8164 [nucl-th]} \BibitemShut {NoStop}%
\bibitem [{\citenamefont {Lin}\ \emph {et~al.}(2005)\citenamefont {Lin}, \citenamefont {Ko}, \citenamefont {Li}, \citenamefont {Zhang},\ and\ \citenamefont {Pal}}]{Lin:2004en}%
  \BibitemOpen
  \bibfield  {author} {\bibinfo {author} {\bibfnamefont {Z.-W.}\ \bibnamefont {Lin}}, \bibinfo {author} {\bibfnamefont {C.~M.}\ \bibnamefont {Ko}}, \bibinfo {author} {\bibfnamefont {B.-A.}\ \bibnamefont {Li}}, \bibinfo {author} {\bibfnamefont {B.}~\bibnamefont {Zhang}},\ and\ \bibinfo {author} {\bibfnamefont {S.}~\bibnamefont {Pal}},\ }\href {https://doi.org/10.1103/PhysRevC.72.064901} {\bibfield  {journal} {\bibinfo  {journal} {Phys. Rev. C}\ }\textbf {\bibinfo {volume} {72}},\ \bibinfo {pages} {064901} (\bibinfo {year} {2005})},\ \Eprint {https://arxiv.org/abs/nucl-th/0411110} {arXiv:nucl-th/0411110} \BibitemShut {NoStop}%
\bibitem [{\citenamefont {Parotto}\ \emph {et~al.}(2020)\citenamefont {Parotto}, \citenamefont {Bluhm}, \citenamefont {Mroczek}, \citenamefont {Nahrgang}, \citenamefont {Noronha-Hostler}, \citenamefont {Rajagopal}, \citenamefont {Ratti}, \citenamefont {Sch\"afer},\ and\ \citenamefont {Stephanov}}]{Parotto:2018pwx}%
  \BibitemOpen
  \bibfield  {author} {\bibinfo {author} {\bibfnamefont {P.}~\bibnamefont {Parotto}}, \bibinfo {author} {\bibfnamefont {M.}~\bibnamefont {Bluhm}}, \bibinfo {author} {\bibfnamefont {D.}~\bibnamefont {Mroczek}}, \bibinfo {author} {\bibfnamefont {M.}~\bibnamefont {Nahrgang}}, \bibinfo {author} {\bibfnamefont {J.}~\bibnamefont {Noronha-Hostler}}, \bibinfo {author} {\bibfnamefont {K.}~\bibnamefont {Rajagopal}}, \bibinfo {author} {\bibfnamefont {C.}~\bibnamefont {Ratti}}, \bibinfo {author} {\bibfnamefont {T.}~\bibnamefont {Sch\"afer}},\ and\ \bibinfo {author} {\bibfnamefont {M.}~\bibnamefont {Stephanov}},\ }\href {https://doi.org/10.1103/PhysRevC.101.034901} {\bibfield  {journal} {\bibinfo  {journal} {Phys. Rev. C}\ }\textbf {\bibinfo {volume} {101}},\ \bibinfo {pages} {034901} (\bibinfo {year} {2020})},\ \Eprint {https://arxiv.org/abs/1805.05249} {arXiv:1805.05249 [hep-ph]} \BibitemShut {NoStop}%
\bibitem [{\citenamefont {Saul’yev}(1957)}]{Saul'yve:1957}%
  \BibitemOpen
  \bibfield  {author} {\bibinfo {author} {\bibfnamefont {V.~K.}\ \bibnamefont {Saul’yev}},\ }\href@noop {} {\bibfield  {journal} {\bibinfo  {journal} {Dokl. Akad. Nauk SSSR}\ }\textbf {\bibinfo {volume} {115}},\ \bibinfo {pages} {1077} (\bibinfo {year} {1957})}\BibitemShut {NoStop}%
\bibitem [{\citenamefont {Saul’yev}(1964)}]{Saul'yve:1964}%
  \BibitemOpen
  \bibfield  {author} {\bibinfo {author} {\bibfnamefont {V.~K.}\ \bibnamefont {Saul’yev}},\ }\href@noop {} {\emph {\bibinfo {title} {{Integration of Equations of Parabolic Type by the Method of Nets}}}}\ (\bibinfo  {publisher} {{Pergamon Press, Oxford}},\ \bibinfo {year} {1964})\BibitemShut {NoStop}%
\bibitem [{\citenamefont {Yang}\ \emph {et~al.}(2022)\citenamefont {Yang}, \citenamefont {Li}, \citenamefont {Lee}, \citenamefont {Lee}, \citenamefont {Kwak}, \citenamefont {Hwang}, \citenamefont {Xin},\ and\ \citenamefont {Kim}}]{yang2022explicit}%
  \BibitemOpen
  \bibfield  {author} {\bibinfo {author} {\bibfnamefont {J.}~\bibnamefont {Yang}}, \bibinfo {author} {\bibfnamefont {Y.}~\bibnamefont {Li}}, \bibinfo {author} {\bibfnamefont {C.}~\bibnamefont {Lee}}, \bibinfo {author} {\bibfnamefont {H.~G.}\ \bibnamefont {Lee}}, \bibinfo {author} {\bibfnamefont {S.}~\bibnamefont {Kwak}}, \bibinfo {author} {\bibfnamefont {Y.}~\bibnamefont {Hwang}}, \bibinfo {author} {\bibfnamefont {X.}~\bibnamefont {Xin}},\ and\ \bibinfo {author} {\bibfnamefont {J.}~\bibnamefont {Kim}},\ }\href@noop {} {\bibfield  {journal} {\bibinfo  {journal} {International Journal of Mechanical Sciences}\ }\textbf {\bibinfo {volume} {217}},\ \bibinfo {pages} {106985} (\bibinfo {year} {2022})}\BibitemShut {NoStop}%
\bibitem [{\citenamefont {Cooper}\ and\ \citenamefont {Frye}(1974)}]{Cooper:1974mv}%
  \BibitemOpen
  \bibfield  {author} {\bibinfo {author} {\bibfnamefont {F.}~\bibnamefont {Cooper}}\ and\ \bibinfo {author} {\bibfnamefont {G.}~\bibnamefont {Frye}},\ }\href {https://doi.org/10.1103/PhysRevD.10.186} {\bibfield  {journal} {\bibinfo  {journal} {Phys. Rev. D}\ }\textbf {\bibinfo {volume} {10}},\ \bibinfo {pages} {186} (\bibinfo {year} {1974})}\BibitemShut {NoStop}%
\bibitem [{\citenamefont {Chattopadhyay}\ and\ \citenamefont {Pal}(2018)}]{Chattopadhyay:2018dth}%
  \BibitemOpen
  \bibfield  {author} {\bibinfo {author} {\bibfnamefont {C.}~\bibnamefont {Chattopadhyay}}\ and\ \bibinfo {author} {\bibfnamefont {S.}~\bibnamefont {Pal}},\ }\href {https://doi.org/10.1103/PhysRevC.98.034911} {\bibfield  {journal} {\bibinfo  {journal} {Phys. Rev. C}\ }\textbf {\bibinfo {volume} {98}},\ \bibinfo {pages} {034911} (\bibinfo {year} {2018})},\ \Eprint {https://arxiv.org/abs/1807.03522} {arXiv:1807.03522 [nucl-th]} \BibitemShut {NoStop}%
\bibitem [{\citenamefont {Pradeep}\ and\ \citenamefont {Stephanov}(2023)}]{Pradeep:2022eil}%
  \BibitemOpen
  \bibfield  {author} {\bibinfo {author} {\bibfnamefont {M.~S.}\ \bibnamefont {Pradeep}}\ and\ \bibinfo {author} {\bibfnamefont {M.}~\bibnamefont {Stephanov}},\ }\href {https://doi.org/10.1103/PhysRevLett.130.162301} {\bibfield  {journal} {\bibinfo  {journal} {Phys. Rev. Lett.}\ }\textbf {\bibinfo {volume} {130}},\ \bibinfo {pages} {162301} (\bibinfo {year} {2023})},\ \Eprint {https://arxiv.org/abs/2211.09142} {arXiv:2211.09142 [hep-ph]} \BibitemShut {NoStop}%
\bibitem [{\citenamefont {{Abbasi}}\ and\ \citenamefont {{Rischke}}(2024)}]{2024arXiv241007929A}%
  \BibitemOpen
  \bibfield  {author} {\bibinfo {author} {\bibfnamefont {N.}~\bibnamefont {{Abbasi}}}\ and\ \bibinfo {author} {\bibfnamefont {D.~H.}\ \bibnamefont {{Rischke}}},\ }\href {https://doi.org/10.48550/arXiv.2410.07929} {\bibfield  {journal} {\bibinfo  {journal} {arXiv e-prints}\ ,\ \bibinfo {eid} {arXiv:2410.07929}} (\bibinfo {year} {2024})},\ \Eprint {https://arxiv.org/abs/2410.07929} {arXiv:2410.07929 [hep-th]} \BibitemShut {NoStop}%
\bibitem [{\citenamefont {Cheng}\ \emph {et~al.}(2009)\citenamefont {Cheng} \emph {et~al.}}]{Cheng:2008zh}%
  \BibitemOpen
  \bibfield  {author} {\bibinfo {author} {\bibfnamefont {M.}~\bibnamefont {Cheng}} \emph {et~al.},\ }\href {https://doi.org/10.1103/PhysRevD.79.074505} {\bibfield  {journal} {\bibinfo  {journal} {Phys. Rev. D}\ }\textbf {\bibinfo {volume} {79}},\ \bibinfo {pages} {074505} (\bibinfo {year} {2009})},\ \Eprint {https://arxiv.org/abs/0811.1006} {arXiv:0811.1006 [hep-lat]} \BibitemShut {NoStop}%
\bibitem [{\citenamefont {Bazavov}\ \emph {et~al.}(2017)\citenamefont {Bazavov} \emph {et~al.}}]{Bazavov:2017dus}%
  \BibitemOpen
  \bibfield  {author} {\bibinfo {author} {\bibfnamefont {A.}~\bibnamefont {Bazavov}} \emph {et~al.},\ }\href {https://doi.org/10.1103/PhysRevD.95.054504} {\bibfield  {journal} {\bibinfo  {journal} {Phys. Rev. D}\ }\textbf {\bibinfo {volume} {95}},\ \bibinfo {pages} {054504} (\bibinfo {year} {2017})},\ \Eprint {https://arxiv.org/abs/1701.04325} {arXiv:1701.04325 [hep-lat]} \BibitemShut {NoStop}%
\bibitem [{\citenamefont {Tsypin}(1994)}]{Tsypin:1994nh}%
  \BibitemOpen
  \bibfield  {author} {\bibinfo {author} {\bibfnamefont {M.~M.}\ \bibnamefont {Tsypin}},\ }\href {https://doi.org/10.1103/PhysRevLett.73.2015} {\bibfield  {journal} {\bibinfo  {journal} {Phys. Rev. Lett.}\ }\textbf {\bibinfo {volume} {73}},\ \bibinfo {pages} {2015} (\bibinfo {year} {1994})}\BibitemShut {NoStop}%
\bibitem [{\citenamefont {Wu}\ and\ \citenamefont {Song}(2019)}]{Wu:2019qfz}%
  \BibitemOpen
  \bibfield  {author} {\bibinfo {author} {\bibfnamefont {S.}~\bibnamefont {Wu}}\ and\ \bibinfo {author} {\bibfnamefont {H.}~\bibnamefont {Song}},\ }\href {https://doi.org/10.1088/1674-1137/43/8/084103} {\bibfield  {journal} {\bibinfo  {journal} {Chin. Phys. C}\ }\textbf {\bibinfo {volume} {43}},\ \bibinfo {pages} {084103} (\bibinfo {year} {2019})},\ \Eprint {https://arxiv.org/abs/1903.06075} {arXiv:1903.06075 [nucl-th]} \BibitemShut {NoStop}%
\end{thebibliography}%
\end{document}